\documentclass[journal]{IEEEtran}
\usepackage[left=0.75in, right=0.75in, top=1in, bottom=0.9in]{geometry}
\usepackage{mathrsfs,amsthm}
\usepackage{cite}
\makeatletter
\def\ps@headings{%
\def\@oddhead{\mbox{}\scriptsize\rightmark \hfil \thepage}%
\def\@evenhead{\scriptsize\thepage \hfil \leftmark\mbox{}}%
\def\@oddfoot{}%
\def\@evenfoot{}}
\makeatother 
\usepackage{graphicx,amsmath,amssymb,stfloats,subfigure,multicol}%amsmath
\usepackage{epsfig,epsf,psfrag,amssymb,amsfonts,latexsym,graphicx,mathrsfs,subfigure}
\usepackage{bbm}
\usepackage{chngpage}
\usepackage[dvips]{color}
\usepackage{textcomp}
\usepackage{hyperref}
\usepackage{url}
\usepackage[hyphenbreaks]{breakurl}
\usepackage[normalem]{ulem}
\usepackage{algpseudocode}
\newsavebox{\ieeealgbox}

\usepackage{array}
\newtheorem{theorem}{Theorem}
\newtheorem{proposition}{Proposition}

\newtheorem{definition}{Definition}

\newcommand*{\QED}{\hfill\ensuremath{\square}}

 \def\old#1{}    % Please don't remove this... This command includes the text to be deleted.

\def\ED{{\mbox{\tiny ED}}}
\def\RED{{\mbox{\tiny R-ED}}}
\def\TLMP{{\mbox{\tiny TLMP}}}
\def\RTLMP{{\mbox{\tiny R-TLMP}}}
\def\LMP{{\mbox{\tiny LMP}}}
\def\RLMP{{\mbox{\tiny R-LMP}}}

\def\nn{\nonumber}
\def\beq{\begin{equation}}
\def\eeq{\end{equation}}
\def\bea{\begin{eqnarray}}
\def\eea{\end{eqnarray}}
\def\ba{\begin{array}}
\def\ea{\end{array}}

\def\bitem{\begin{itemize}}
\def\eitem{\end{itemize}}
\def\ben{\begin{enumerate}}
\def\een{\end{enumerate}}

\def\eg{{\it e.g., \/}}

\def\ie{{\it i.e.,\ \/}}

\definecolor{bgrd}{rgb}{1,1,1}
\definecolor{gray}{rgb}{0.5,0.5,0.5}
\definecolor{dkr}{rgb}{0.7,0.1,0.2}
\definecolor{dkb}{rgb}{0.1,0.1,0.8}

\makeatletter
\newdimen{\captionwidth}
\long\def\@makecaption#1#2{%
\captionwidth .9\hsize% use current value of \hsize
\vskip 10pt%
\setbox\@tempboxa\hbox{#1: #2}%
  \ifdim \wd\@tempboxa >\captionwidth%
    \setbox\@tempboxa\hbox{#1:\hspace*{.5em}}%
    \hfil\parbox{\captionwidth}{\raggedright\hangindent \wd\@tempboxa%
    \hangafter=1\unhbox\@tempboxa#2}\hfill%
  \else\centerline{\box\@tempboxa}%
  \fi
}
\makeatother
\def\scalefig#1{\epsfxsize #1\textwidth}

\def\edoc{\end{document}}

\def\T{\mbox{\tiny T}}

\newcommand{\Hmsc}{\mathscr{H}}

\def\zetabf{\hbox{\boldmath$\zeta$\unboldmath}}
\def\etabf{\hbox{\boldmath$\eta$\unboldmath}}

\def\lambdabf{\hbox{\boldmath$\lambda$\unboldmath}}
\def\mubf{\hbox{\boldmath$\mu$\unboldmath}}

\def\pibf{\hbox{\boldmath$\pi$\unboldmath}}
\def\rhobf{\hbox{\boldmath$\rho$\unboldmath}}

\def\Pibf{{\bf \Pi}}
\def\thetabf{{\mbox{\boldmath$\theta$\unboldmath}}}
\def\dbf{{\bf d}}

\def\fbf{{\bf f}}
\def\gbf{{\bf g}}

\def\pbf{{\bf p}}
\def\qbf{{\bf q}}
\def\rbf{{\bf r}}

\def\xbf{{\bf x}}

\def\rbf{{\bf r}}
\def\xbf{{\bf x}}

\def\Abf{{\bf A}}

\def\Gbf{{\bf G}}

\def\Xbf{{\bf X}}

\def\Gc{{\cal G}}
\def\Hc{{\cal H}}

\def\Lc{{\cal L}}

\def\Pc{{\cal P}}

\begin{document}

\title{Pricing Multi-Interval Dispatch under Uncertainty\\
Part I: Dispatch-Following Incentives}
\author{Ye Guo,~\IEEEmembership{Senior Member,~IEEE,}
		Cong Chen,~\IEEEmembership{Student Member,~IEEE,}
		and~Lang~Tong,~\IEEEmembership{Fellow,~IEEE}% <-this % stops a space
\thanks{\scriptsize Part of the work was presented at the 56th Allerton Conference on Communication, Control, and Computing \cite{Guo&Tong:18Allerton} and 2019 IEEE PESGM \cite{Tong:19PESGM}.}
\thanks{\scriptsize
Ye Guo (\url{guo-ye@sz.tsinghua.edu.cn}) is with Tsinghua Berkeley Shenzhen Institute, Shenzhen, P.R. China.  Cong Chen and Lang Tong (\{cc2662,lt35\}@cornell.edu) are with the School of Electrical and Computer Engineering, Cornell University,  USA.  Corresponding authors: Lang Tong and Ye Guo.}
\thanks{\scriptsize The work of L. Tong and C. Chen is supported in part by the National Science Foundation under
Award 1809830 and 1932501, Power Systems and Engineering Research Center (PSERC) Research Project M-39. The work of Y. Guo is supported in part of the National Science Foundation of China under Award 51977115.}}

\maketitle

\begin{abstract}
Pricing multi-interval economic dispatch of electric power under operational uncertainty is considered in this two-part paper. Part I investigates dispatch-following incentives of profit-maximizing generators and shows that, under mild conditions, no uniform-pricing scheme for the rolling-window economic dispatch provides dispatch-following incentives that avoid discriminative out-of-the-market uplifts.
A nonuniform pricing mechanism, referred to as the temporal locational marginal pricing (TLMP), is proposed.  As an extension of the standard locational marginal pricing (LMP), TLMP takes into account both generation and ramping-induced opportunity costs.  It eliminates the need for the out-of-the-market uplifts and guarantees full dispatch-following incentives regardless of the accuracy of the demand forecasts used in the dispatch.    It is also shown that, under TLMP, a price-taking market participant has incentives to bid truthfully with its marginal cost of generation.   Part II of the paper extends the theoretical results developed in Part I to more general network settings.  It investigates a broader set of performance measures, including the incentives of the truthful revelation of ramping limits, revenue adequacy of the operator, consumer payments, generator profits, and price volatility under the rolling-window dispatch model with demand forecast errors.
\end{abstract}


\section{Introduction} \label{sec:intro}

We consider the problem of pricing multi-interval look-ahead economic dispatch when generators are ramp-constrained and demand forecasts inaccurate.   This work is motivated by  recent discussions among system operators on the need for ramping products in response to the ``duck-curve'' effect of renewable integrations \cite{Xie&Luo&Obadina:11PESGM,Peng&Chatterjee:13FERC,CAISO_FRP:15,Mickey:15,Parker:15,%
Schiro:17}.   A well-designed multi-interval look-ahead dispatch that anticipates trends of future demand can minimize the use of more expensive ramp resources. %\cite{XIA&Elaiw:10EPSR}.

A standard implementation of a look-ahead dispatch is the so-called {\em rolling-window dispatch}, where the operator optimizes the dispatch over a few scheduling intervals into the future based on load forecasts.  The dispatch for the immediate scheduling interval ({\it a.k.a. the binding interval}) is implemented while the dispatch for the subsequent intervals serves as an advisory signal and is updated sequentially.  A common practice to price the rolling-window dispatch is the rolling-window version of the multi-interval locational marginal pricing (LMP).

LMP is a uniform pricing mechanism across generators and demands at the same location in the same scheduling interval. For the single-interval pricing problem, LMP has remarkable properties.  LMP supports an efficient market equilibrium such that a profit-maximizing generator has no incentive to deviate from the central dispatch.   For a competitive market with a large number of generators, a price-taking generator has the incentive to bid truthfully at its marginal cost of generation.    LMP also guarantees a nonnegative merchandising surplus for the system operator. As a uniform pricing scheme, LMP is transparent to all market participants, and the price can be computed easily as a by-product of the underlying economic dispatch.

Most of the attractive features of LMP are lost, unfortunately, when the  rolling-window version of LMP (R-LMP) is used and demand forecasts inaccurate.  Indeed, even if perfect forecasts are used in R-LMP,  many nice properties of LMP  are not guaranteed.  In particular, a missing-money scenario arises when a  generator is asked to hold back its generation in order to provide ramping support for the system to meet  demands in future intervals.  In doing so, the generator incurs an opportunity cost and may be paid below its  offered price to generate.   Expecting compensations in future intervals for the opportunity costs, the generator disappoints when the anticipated higher payments do not realize due to changing demand forecasts.    Examples of such scenarios are well known and also illustrated in Example 2 in Sec~\ref{sec:example}.  It turns out that such examples are not isolated instances unique to R-LMP.  As we show in  Theorem~\ref{thm:UniformPricingLOC} in Sec.~\ref{sec:incentives}, they occur under all uniform pricing schemes.

To ensure that generators are adequately compensated, the operator  provides the so-called {\em uplift payments}  to generators suffering from underpayments in an {\em out-of-the-market settlement.}    The roles of  uplifts have been discussed extensively in the literature \cite{Gribik&Hogan&Pope:07,Zhang&Luh&Litvinov&Zheng&Zhao:09PESGM,Al-Abdullah&Abdi-Khorsand&Hedman:15TPS,Zhang&Hedman:19NAPS}.   Such settlements are typically discriminative and subject to manipulation.   Examples exist that, under LMP,  a price-taking generator may have incentives to deviate from truthful-bidding to take advantage the out-of-the-market settlements.  See Appendix L.

\subsection{Related work}
Wilson discussed the issue of pricing distortion introduced by ramping in \cite{Wilson:02E}.  He pointed out that the cause of such pricing distortion is that the optimization model used in price formation is imperfect.  In the rolling-widow dispatch context, both the imperfection of demand forecasts and the limited look-ahead distort the dispatch-following incentives.  The use of out-of-the-market uplifts further distorts the truthful-bidding incentives.  The dispatch-following incentive issues in pricing multi-interval dispatch  have been widely discussed in the literature \cite{Peng&Chatterjee:13FERC,Ela&OMalley:16TPS,%
Schiro:17,Guo&Tong:18Allerton,Hua&etal:19TPS,Zhao&Zheng&Litvinov:19arxiv,Zhang&Hedman:19NAPS}, although a formal way of analyzing such issues is lacking.  The effects of the out-of-the-market uplifts on truthful-bidding incentives are not well understood.

Several marginal cost pricing schemes have been proposed for the rolling-window dispatch policies.  The flexible ramping product (FRP) \cite{CAISO_FRP:15} treats ramping as a product to be procured and priced uniformly as part of the real-time dispatch.  FRP is a two-part tariff consisting of prices of energy and ramping.  Ela and O'Malley proposed the cross-interval marginal price (CIMP) in \cite{Ela&OMalley:16TPS} defined by the sum of marginal costs with respect to the demands in the binding and the future (advisory) intervals.  Multi-settlement pricing schemes are proposed in \cite{Schiro:17FERC,Zhao&Zheng&Litvinov:19arxiv} that generalize the existing two-settlement day-ahead and real-time markets.

Deviating from marginal cost pricing are two recent proposals aimed at minimizing the out-of-the-market payments; both employ separate pricing optimizations that are different from that used in the economic dispatch.    The price-preserving multi-interval pricing (PMP), initially suggested by Hogan in \cite{Hogan:16} and formalized in \cite{Hua&etal:19TPS}, adds to the objective function the loss-of-opportunity cost for the generators for the realized prices and dispatch decisions.   In contrast, the constraint-preserving multi-interval pricing (CMP) proposed in  \cite{Hua&etal:19TPS} fixes the past dispatch decisions and penalizes ramping violations.  Both have shown improvements over the standard R-LMP policy.

 All existing pricing schemes for multi-interval economic dispatch are based on uniform pricing mechanisms.  To our best knowledge, no existing pricing policies can provide dispatch-following incentives that eliminate discriminative out-of-the-market settlements.

\subsection{Summary of results, contexts, and limitations}
The main contribution of this work is threefold.   First, we show in Theorem~\ref{thm:UniformPricingLOC} that price discrimination is  unavoidable in pricing rolling-window dispatch.  Specifically,   all uniform pricing mechanisms require some level of out-of-the-market uplifts under the rolling-window dispatch model.  While uniform pricing schemes are transparent and  non discriminative within the market clearing process, it is the  out-of-the-market uplift payments that make the overall payment scheme discriminative.

Second, we  generalize LMP to a nonuniform pricing scheme, referred to as the {\em temporal locational marginal pricing (TLMP)}.  TLMP prices the production of a generator $i$ based on its contribution to meeting the demand in interval $t$.  In doing so, TLMP encapsulates both generation and ramping-induced opportunity costs in each interval.

As shown in Proposition~\ref{prop:TLMP}, TLMP decomposes into energy and ramping prices:
\beq \label{eq:TLMP0}
\pi_{it}^{\TLMP}=\pi_{t}^{\LMP} + \mu_{it}-\mu_{i(t-1)},
\eeq
where $\pi_{t}^{\LMP} $ is the standard LMP, and the second term is the increment of the Lagrange multipliers associated with the ramping constraints in the economic dispatch optimization, from $\mu_{i(t-1)}$  in interval $(t-1)$ to $\mu_t$ in interval $t$.   The above decomposition is analogous to the energy-congestion price decomposition of LMP.  TLMP naturally reduces to LMP in the absence of binding ramping constraints.

Third, we establish several key properties that make TLMP a viable and potentially attractive alternative to standard uniform pricing schemes. A key  property of TLMP is that, under the dispatch and pricing models assumed in this paper, the rolling-window implementation of TLMP (R-TLMP) eliminates the need of out-of-the-market uplifts for the rolling-window economic dispatch under arbitrary forecast errors.  Whereas all pricing schemes are necessarily discriminative,  R-TLMP stands out as one that discriminates inside rather than outside  the market clearing process.  This property ties  real-time pricing closely to the actually realized ramping conditions.

As a generalization of LMP, TLMP extends some of the important properties of LMP to the rolling-window  multi-interval pricing setting, thanks to the property that R-TLMP is a strong equilibrium price that decouples the profit maximization problem over the entire scheduling horizon into  single-interval ones.  A significant property of TLMP (Theorem~\ref{thm:TLMPBid}) is that a price-taking profit-maximizing generator has the incentive to bid truthfully with its marginal cost of generation.   In other words, there is no need for a generator to internalize ramping-induced opportunity  costs.  Such a property, however,  does not hold for the rolling-window implementation of the multi-interval LMP.  See Appendix L.

Also significant (Proposition 3 of Part II) is that, under TLMP, the operator's merchandising surplus is the sum of congestion and ramping surplus, which has significant implications on the revenue adequacy of ISO.    We also demonstrate that, under TLMP, the generators have incentives for truthful  revelation of ramping limits, and there are incentives for the generators to improve their ramping capabilities.

Given that TLMP is discriminatory, one may question how different it is from other discriminative pricing schemes such as the  pay-as-bid (PAB) pricing.   The differences between TLMP and PAB pricing are significant; TLMP is much closer to LMP than it is to PAB.   Comparing with LMP, PAB is more vulnerable to manipulative bidding behaviors, and a market participant has little incentive to bid truthfully.   In contrast, TLMP inherits and extends (in Theorem \ref{thm:TLMPBid}) the property of LMP (under the single-interval model) that a price taker generator bids truthfully with its marginal cost.

Discriminative pricing is often criticized for its lack of  transparency, which makes it difficult for the operator to provide public pricing signals to market participants.  Because of the decomposition of TLMP into the uniform energy price (LMP)  and a discriminative ramping price in (\ref{eq:TLMP0}), the energy part of TLMP (LMP) is transparent to all participants.   The ramping price part of TLMP, like the out-of-the-market uplifts, is nontransparent and discriminative.  In this aspect, TLMP has the same level of transparency as in LMP, although the amount of the discriminative payments under TLMP and  uniform prices can be quite different.  See Part II of this paper for a numerical comparison \cite{Chen&Guo&Tong:20TPS}.

Finally, in Part II of the paper, we generalize the theory of dispatch-following incentives to more general models that include network constraints and discuss a broader set of incentive and performance issues through numerical simulations.    When comparing different pricing schemes, our results shine lights on practical tradeoffs along several dimensions:  the revenue adequacy of the ISO, consumer payments, generator profits, and price volatilities.

A few words are in order on the scope and limitations of this paper.
We do not model strategic behaviors of the generators, nor do we consider those market models that the market operator does not price ramping costs and lets the generators internalize their individual ramping costs. We discuss in Sec.~\ref{sec:discussion} some of the implications of these omissions.   We also ignore the role of unit commitment and the costs of  reserves.   In Part I,  we illustrate the properties of LMP and TLMP with a toy example.  Generalizations to systems with network constraints and more elaborate numerical examples are in Part II.

\subsection{Notations and nomenclature}
Designated symbols are listed in Table~I.
Otherwise, notations used here are standard.  We use $(x_1,\cdots,x_N)$ for a {\em column vector} and  $[x_1,\cdots,x_N]$ a {\em row vector.}    All vectors are denoted by lower-case boldface letters,  nominally as columns.   The transpose of vector $\xbf$ is denoted by $\xbf^{\tiny \intercal}$.  Matrices are boldface capital letters. Matrix $\Xbf=[x_{ij}]$  is a matrix with $x_{ij}$ as its $(i,j)$th entry.   Similar to the vector notation, matrix $\Xbf=[\xbf_1,\cdots,\xbf_N]$ has  $\xbf_i$ as its $i$th column, and matrix $\Xbf=(\xbf^{\intercal}_1,\cdots,\xbf^{\intercal}_N)$ has $\xbf_i^{\intercal}$ as its $i$th row.

{\small
\begin{table}[h]\label{tab:symbols}
\caption{\small Major symbols (in alphabetic order).}
\begin{center}
\begin{tabular}{|ll|}
\hline
${\bf 0}, {\bf 1}$: & vector of all zeros and ones.\\
$\Abf$: & a $W\times W$ {\em lower bi-digonal matrix} with 1 on\\
&the diagonals and -1 on the off diagonals.\\\hline
% the upper bi-diagonal matrix with $-1$ as \\
% & diagonals and 1 as off diagonals.  \\\hline
$d_t$: &  the demand in interval $t$.  \\
$\dbf_t$:  &$\dbf_t=(d_t,\cdots, d_{t+W-1})$,  the demand in the \\
& look-ahead window of $W$ intervals.\\
$\dbf$: & $\dbf=(d_1,\cdots, d_T)$   the overall demand vector. \\
$\hat{d}_t, \hat{\dbf}_t$: & demand forecasts of $d_t$ and $\dbf_t$.\\\hline
$f_{it}(\cdot)$: & the bid-in cost  of generator $i$ in interval $t$. \\
$F(\cdot)$: & aggregated bid-in cost curve in $W$ or $T$ intervals.\\
$F_{-it}(\cdot)$: & aggregated bid-in cost curve excluding generation \\
& from generator $i$ in interval $t$.\\\hline
$g_{it}$: & generation/dispatch of generator $i$ in interval $t$ \\
$\gbf[t]$: & generation/dispatch vector for all generators in\\
 & interval $t$, $\gbf[t]=(g_{1t},\cdots, g_{Nt})$.\\
$\Gbf$: & generation/dispatch matrix. $\Gbf=\big[\gbf[1],\cdots,\gbf[T]\big]$.\\
$\gbf_i$: & dispatch of generator $i$ over a scheduling window, \\
& \eg $\gbf_i=(g_{i1},\cdots, g_{iW})$ or $\gbf_i=(g_{i1},\cdots, g_{iT})$. \\\hline
$\Gc_t$: & the  look-ahead dispatch policy at time $t$.\\
$\Gc^{\mbox{\tiny ED}}_t$: & the look-ahead  economic dispatch policy at time $t$.\\
$\gbf_i^{\ED}$: &  one-shot economic dispatch for generator $i$.\\
$\gbf_i^{\RED}$: & rolling-window economic dispatch for generator $i$.\\\hline
$\Hmsc_t$: & scheduling window $\Hmsc_t=\{t,\cdots, t+W-1\}$.\\
$\Hmsc$: & scheduling horizon  $\Hmsc=\{1,\cdots, T\}$.\\\hline
$\mbox{LOC}$: & lost-of-opportunity cost uplift. \\
$\mbox{MW}$: & make-whole uplift.\\\hline
$\Pc_t, \Pc^{\mbox{\tiny LMP}}_t$: & multi-interval pricing policy and LMP pricing policy.\\
$\pibf^{\LMP},\pibf^{\TLMP}$: & one-shot LMP and TLMP.\\
$\pibf^{\RLMP},\pibf^{\RTLMP}$: & rolling-window LMP/TLMP. \\\hline
$q_{it}(\cdot), \qbf_i(\cdot)$: & true cost of generation of generator $i$.\\
$T$: & total number of scheduling intervals. \\
$W$:& scheduling window size. $W\le T$.\\\hline
\end{tabular}
%\caption{\small Major symbols (in alphabetic order).}
\end{center}
\end{table}
}

\section{Multi-interval dispatch and pricing models} \label{sec:model}
We consider a {\em bid-based real-time electricity market}  involving one inelastic demand,  $N$ generators, and a system (market) operator.  The scheduling period of generations involves $T$ unit-length  intervals $\Hmsc=\{1,\cdots, T\}$, where interval $t$ covers the time interval $[t,t+1)$.  Typically,   $T$  is the number of intervals in a day.

We assume that each generator produces a generation offer that includes a bid-in cost curve along with its generation and ramping limits.
The operator collects bids from all generating firms, allocates generation levels to all generators in the form of dispatch signals, and determines the prices of electricity in each scheduling interval. We assume that, in pricing multi-interval dispatch, the operator incorporates generation and ramping constraints.   Because the bid of a generator  represents  its willingness to generate, the generator expects the total payment received over $T$ intervals to be no less than that computed from its offered prices;   anything less needs to  be compensated  by some forms of uplift payments outside the market clearing process.

Part I of the paper assumes a  single-bus network, which is generalized in Part II to networks with $M$ buses subject to network constraints.
 We introduce two multi-interval scheduling and pricing models.  One is the {\em one-shot model} that sets generation dispatch and prices over the entire scheduling period at once, the other the {\em rolling-window model} that sets the dispatch levels and prices sequentially with demand forecasts for several intervals into the future.

\subsection{One-shot multi-interval dispatch and pricing policies}
 At $t=1$, the operator obtains the demand forecast vector
 $\hat{\dbf}=(\hat{d}_{1},\cdots, \hat{d}_{T})$ over the entire scheduling horizon $\Hmsc$, where $\hat{d}_{t}$ is the  demand forecast for  interval $t$.    Let the actual demand be $\dbf=(d_1,\cdots, d_T)$. We assume that the forecast of the first interval is accurate, \ie $\hat{d}_{1}=d_1$.

A {\em one-shot dispatch} schedules generations  over the $T$-interval scheduling horizon $\Hmsc$ based on the initial forecast $\hat{\dbf}$.  Let $g_{it}$ be the dispatch of generator  $i$ in interval $t$, $\gbf_i=(g_{i1},\cdots, g_{iT})$ the dispatch  for generator $i$ over $\Hmsc$, $\gbf[t]=(g_{1t},\cdots, g_{Nt})$ the dispatch for all generators in interval $t$, and the $N\times T$ matrix $\Gbf = \big[\gbf[1],\cdots, \gbf[T]\big]$  the dispatch matrix  with $\gbf_i^{\intercal}$ as its $i$th row.

A {\em one-shot dispatch policy} $\Gc$  maps the demand forecast $\hat{\dbf}$ and the initial generation $\gbf[0]$ to a { dispatch matrix} $\Gbf$:
\[
\Gc(\hat{\dbf},\gbf[0]) = \Gbf,
\]
where $\gbf[0]$ imposes the initial ramping constraints on the generations  in the first interval.

Similarly, a {\em one-shot pricing policy} $\Pc$  sets the prices  in all intervals at once.   A one-shot uniform price is defined by a  vector $\pibf=(\pi_1,\cdots,\pi_T)$ with $\pi_t$ being the price of electricity in interval $t$ for all generators and the demand.  For a nonuniform pricing policy, $\Pc$ sets  $\pibf_0$ the price vector for the demand and  $\pibf_i=(\pi_{i1},\cdots, \pi_{iT})$ for generator $i$, for $i=1,\cdots, N$.

\subsection{One-shot economic dispatch and LMP}
A special case of the one-shot dispatch is the
{\em  multi-interval economic dispatch} $\Gc^{\rm\tiny ED}$  over $\Hmsc$.   Let the aggregated bid-in cost function be
\beq \label{eq:G}
F(\Gbf) := \sum\limits_{i=1}^N\sum\limits_{t\in \Hmsc} f_{it}(g_{it}),
\eeq
where $f_{it}(\cdot)$ is the bid-in cost curve\footnote{The derivative of the bid-in cost curve represents the supply curve of the generator.}  of generator $i$  in interval $t$, assumed to be convex and almost everywhere differentiable for all $t$ and $i$ throughout the paper.  Note that $f_{it}(\cdot)$  is not necessarily equal to the actual generation cost $q_{it}(\cdot)$.

The dispatch policy $\Gc^{\rm\tiny ED}$ is defined by
\beq \label{eq:ED}
\begin{array}{lrl}
\Gc^{\tiny\rm ED}: &\underset{\{\Gbf=[g_{it}]\}}{\rm minimize}   &  F(\Gbf)  \\
&  \mbox{subject to} & \mbox{for all $i$ and  $t \in \Hmsc$}\\
& \lambda_{t}: & \sum \limits_{i=1}^N g_{it}= \hat{d}_{t}, \\
& (\underline{\mu}_{it},\bar{\mu}_{it}):  &  -\underline{r}_i\le g_{i(t+1)}-g_{it} \le \bar{r}_{i}\\
& & \hfill 0 \le  t \le T-1,\\
& (\underline{\rho}_{it},\bar{\rho}_{it}):   & 0 \le g_{it} \le \bar{g}_{i}, \\
\end{array}
\eeq
where $\bar{g}_{i}$  the generation capacity, and $(\underline{r}_i,\bar{r}_{i})$ the down and up ramp-limits, $\lambda_t$ the dual variable for the equality constraints, and  $(\underline{\rho}_{it}, \bar{\rho}_{it},\underline{\mu}_{it},\bar{\mu}_{it}) \ge {\bf 0}$ are dual variables for the inequality constraints\footnote{Throughout the paper, all inequalities are written in the form of $v(x)\le 0$ with a non-negative dual variable.}.

The {\em one-shot locational  marginal price\footnote{We retain the LMP terminology even though the model considered here does not involve a network.}}  (LMP for short) is a uniform price $\pibf^{\mbox{\tiny LMP}}=(\pi^{\mbox{\tiny LMP}}_t)$ with $\pi^{\mbox{\tiny LMP}}_t$ defined by  the marginal cost
of  generation with respect to the demand in interval $t$.     In particular, we have, by the envelope theorem,
\[
\pi^{\mbox{\tiny LMP}}_t := \frac{\partial}{\partial \hat{d}_t} F(\Gbf^{\ED}) = \lambda^*_t,~~t=1,\cdots, T,
\]
where $\Gbf^{\ED}$ and $\lambda^*_t$ are part of a solution to (\ref{eq:ED}).

\subsection{Rolling-window look-ahead dispatch model}
\begin{figure}[h]
\center
\begin{psfrags}
\psfrag{G1}[l]{\normalsize $\Gc_{1}$}
\psfrag{G2}[l]{\normalsize $\Gc_{2}$}
\psfrag{G3}[l]{\normalsize $\Gc_{3}$}
\psfrag{t=1}[c]{\small $t=1$}
\psfrag{t=2}[c]{\small $t=2$}
\psfrag{t=3}[c]{\small $t=3$}
\scalefig{0.38}\epsfbox{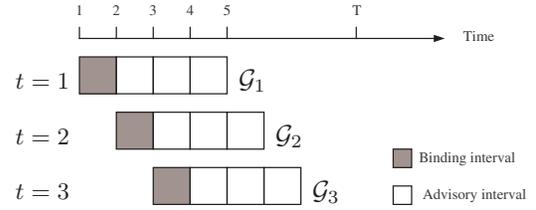}
\end{psfrags}
\caption{\small Rolling-window dispatch with window size $W=4$ generated from one-shot dispatch policy $\Gc_t$. The same applies also to the rolling-window pricing.}
\label{fig:roll}
\end{figure}

A rolling-window dispatch policy $\Gc=(\Gc_1,\cdots, \Gc_T)$ is defined by a sequence of $W$-interval look-ahead policies that  generate dispatch signals $\gbf[1],\cdots, \gbf[T]$ sequentially, as illustrated in Fig.~\ref{fig:roll}. At time $t$, the policy $\Gc_t$ has a look-ahead {\em scheduling window} of $W$ intervals, denoted by $\Hmsc_t =\{t,\cdots, t+W-1\}$.       The interval $t$ is called the {\em binding interval} and the rest of  $\Hmsc_t$  the {\em advisory intervals.}  As time $t$ increases, $\Hmsc_t$  slides across the entire scheduling period $\Hmsc$.

At time $t$, a $W$-interval one-shot policy $\Gc_t$ maps demand forecast $\hat{\dbf}_t=(\hat{d}_{t},\cdots, \hat{d}_{t+W-1})$ and previously realized generation $\gbf[t-1]$ to an $N\times W$  generation scheduling matrix  $\hat{\Gbf}_t$ over $\Hmsc_t$:
\[
\Gc_{t}(\hat{\dbf}_t,\gbf[t-1])=\big[\hat{\gbf}[t],\cdots, \hat{\gbf}[t+W-1]\big] = \hat{\Gbf}_t.
\]
The rolling window policy $\Gc$ sets generation in interval $t$ by $\gbf[t]:=\hat{\gbf}[t]$.  The rest of columns of $\hat{\Gbf}_t$ are not implemented.

Similarly, a {\em rolling-window pricing policy} $\Pc$ is defined by a sequence one-shot pricing policies $(\Pc_1,\cdots, \Pc_T)$.  At time $t$, $\Pc_t$   sets  the prices over $\Hmsc_t$, and  the price in the binding interval $t$ is implemented by $\Pc$.

As an example, the rolling-window economic dispatch policy  $\Gc^{\RED} = (\Gc^{\ED}_1,\cdots,\Gc_T^{\ED})$ where $\Gc_t^{\ED}$ is the $W$-window one-shot economic dispatch defined in (\ref{eq:ED}) with $T=W$ and $\hat{\dbf}=\hat{\dbf}_t$.  The rolling-window LMP policy $\Pc^{\RLMP}$  is defined by a sequence of  $W$-interval LMP policies $(\Pc_1^{\LMP},\cdots, \Pc_T^{\LMP})$.

%A {\em one-shot} policy covers the  the entire scheduling horizon, \ie  $W=T$ and $t=1$. For such policies, we drop the subscript in their notations.

\section{Dispatch-following incentives and Uplifts} \label{sec:incentives}
We say that a pricing mechanism provides  {\em dispatch-following incentives} if, given the {\em realized prices}, profit-maximizing generators, by themselves, would have produced generations that match the operator's dispatch.  Applying market equilibrium models for dispatch-following incentives, we consider two types of  incentives: (i) the {\em ex-post incentive} that applies to the entire scheduling period $\Hmsc$ after all generations have been realized; (ii) the {\em ex-ante incentive} that applies to only the current (binding) scheduling interval. The former guarantees dispatch-following incentives when a generator considers the total profit over the entire scheduling period.
The latter guarantees dispatch-following incentives only for the binding interval.

\subsection{Ex-post incentives and general equilibrium}
For a multi-interval dispatch and pricing problem, generations and consumptions in each interval  are part of a market separate from those in other intervals;  we thus have  a set of $T$ inter-dependent markets over $\Hmsc$.   For purposes of analyzing dispatch-following incentives, we borrow the notion of general equilibrium \cite[p. 547]{Mas-Colell&Winston&Green:95book}  for the multi-interval pricing problem.

\begin{definition}[General equilibrium] \label{def:geq}
{\em  Let $\dbf$ be the actual demand, $\gbf_i$ the dispatch  for generator $i$ and $\pibf$ the vector of electricity prices over the entire scheduling period $\Hmsc$.  Let the $N\times T$ matrix  $\Gbf=(\gbf_1^{\intercal},\cdots,\gbf_N^{\intercal})$ be the realized generation matrix  for all generators.  We say  $(\Gbf, \pibf)$
forms a} general equilibrium {\em if the following market clearing and individual rationality conditions are satisfied:}
\ben
\item { Market clearing condition:}
\[
\sum_{i=1}^N g_{it}=d_t~~\mbox{for all $t \in \Hmsc$}.
\]
\item { Individual rationality condition:} {\em for all $i$, the dispatch $\gbf^{(i)}=(g_{i1},\cdots, g_{iT})$ is the solution to the individual profit maximization:}
\beq \label{eq:geq}
\begin{array}{ll}
\underset{(g_1,\cdots,g_T)}{\rm maximize} & \sum_{t=1}^T (\pi_{t} g_{t} - f_{it}(g_{t})) \\[0.25em]
 {\rm subject~to}& \mbox{\rm for all $t=0,\cdots,T-1$},\\
  & -\underline{r}_i\leq g_{t+1}-g_{t} \leq \bar{r}_i, \\
  & 0\leq g_{t}\leq\bar{g}_{i}, \forall t \in \Hmsc.
\end{array} \hfill
\eeq
\een
{\em We call $\pibf$ an} equilibrium price supporting {generation $\Gbf$.}
\end{definition}

\vspace{0.5em}
In the context of analyzing dispatch-following incentives, we are interested in whether price signal $\pibf$ and dispatch  $\Gbf$ satisfy the general equilibrium condition.  It turns out that, in the absence of forecasting error, the one-shot LMP supports the one-shot economic dispatch as stated  in Theorem~\ref{thm:LMP}.
This result is analogous to the well-known property of LMP \cite{Wu&Variaya&Spiller&Oren:96JRE}.
%The proof is neglected for brevity; it follows immediately from that multi-interval economic dispatch optimization (\ref{eq:ED}) is convex.

\begin{theorem}[LMP as a General Equilibrium Price] \label{thm:LMP} When there is no forecast error, $\hat{\dbf}=\dbf$,  the one-shot economic dispatch matrix  $\Gbf^{\ED}$ and the one-shot LMP  $\pibf^{\LMP}$   form a general equilibrium.
\end{theorem}

  As a general equilibrium price, $\pibf^{\LMP}$ does not guarantee that $\pi^{\LMP}_{it}g_{it} \ge  f_{it}(g_{it})$  for all $(i,t)$. In other words, a generator may be underpaid in some intervals despite that the generator is maximally compensated under $\pibf^{\LMP}$ over the entire scheduling period.   See Example 1 in
Sec.~\ref{sec:example}.

\subsection{Ex-ante incentives and partial equilibrium}
When the  rolling-window dispatch is used, the forecasts in the look-ahead window (hence the dispatch over the window)  change, which creates the  {\em missing payment problem} even when the forecast over the look-ahead window is perfect.

Consider the example of rolling-window economic dispatch $\Gc^{\RED}$ and LMP $\Pc^{\RLMP}$ policies.  Suppose that  a generator $i$ is underpaid in interval $t$, \ie $f_{it}(g_{it}^{\RLMP}) \ge \pi_t^{\RLMP}g_{it}^{\RED}$. Because $g_{it}^{\RED}$ is generated by the $W$-window economic dispatch based on forecast $\hat{\dbf}_t$, generator $i$ expects  the underpayment in interval $t$ be compensated later in $t' \in \Hmsc_t$.
At time $t'$, however, a different forecast $\hat{\dbf}_{t'}$ is used to generate dispatch $g_{it'}^{\RED}$.  There is no guarantee that $\pi_{t'}^{\RLMP}$ is high enough to compensate for the loss incurred in the interval $t$, hence the missing payment problem.

To provide dispatch-following incentives under forecasting uncertainty, we need stronger equilibrium conditions.

\begin{definition}[Partial equilibrium and strong equilibrium]\label{def:peq}
{\em Consider price vector $\pibf=(\pi_1,\cdots, \pi_T)$ and generation matrix $\Gbf$ over the entire scheduling horizon $\Hmsc$.  The dispatch-price pair $(\gbf[t],\pi_t)$ in interval $t$ is a partial equilibrium if it satisfies the market clearing and individual rationality conditions in interval $t$:}
\ben
\item { Market clearing condition:} $\sum_{i=1}^N g_{it} =d_t$;
\item { Individual rationality condition:} {\em for all $i$, the dispatch of signal $g_{it}$ is the solution to the individual profit maximization:}
\beq \label{eq:peq}
\begin{array}{ll}
\underset{g}{\rm maximize} & (\pi_{t} g - f_{it}(g)) \\[0.25em]
 {\rm subject~to}&  0\leq g \leq \bar{g}_{i}\\
%  & -\underline{r}_i\leq g_{i(t+1)}-g \leq \bar{r}_i. \\
    & -\underline{r}_i\leq g-g_{i(t-1)} \leq \bar{r}_i. \\
\end{array} \hfill
\eeq
\een
{\em The dispatch-price pair} $(\Gbf, \pibf)$  {\em is a} strong equilibrium {\em if $(\Gbf,\pibf)$ is a general equilibrium and $(\gbf[t],\pi_t)$ a partial equilibrium for all $t$.}
\end{definition}

The notion of partial equilibrium used here is slightly different from the standard because of the sequential nature of multi-interval dispatch and pricing problems.  At time $t$, the dispatch in the interval $t$ is necessarily constrained by the past dispatch. The dispatch in the future intervals is advisory and subject to change, which is the reason that only the ramping constraints from the previous interval are imposed.

The strong equilibrium conditions impose stricter constraints than that required by the general or partial equilibrium definitions; strong equilibrium implies general equilibrium.  Unlike the case of a general equilibrium price that only needs to satisfy the rationality condition at the end of the scheduling horizon, a strong equilibrium price must provide a dispatch-following incentive in {\em every interval} independent of future realized dispatches.  Consequently, even if schedules and prices may change, for the binding interval, there is no incentive for the generator to deviate from the dispatch signal.

An immediate  corollary of Theorem~\ref{thm:LMP}  is that, in the absence of ramping constraints,  $(\Gbf^{\ED},\pibf^{\LMP})$ forms a strong equilibrium.  However,  we also know from Example 1  in Sec.~\ref{sec:example} that, when ramping constraints are binding,  $(\Gbf^{\ED},\pibf^{\LMP})$  may not be a strong equilibrium. Does there exist a  uniform price $\pibf$ such that $(\Gbf^{\ED},\pibf^{\LMP})$  satisfies the strong equilibrium conditions? %The answer turns out to be negative in general.

\begin{theorem}[Strong equilibrium and uniform price]
Let $\hat{\dbf}=\dbf$ be the actual demand and $\Gbf^{\ED}$ the one-shot economic dispatch over $\Hmsc$.  If  there exists a generator $i$ and an interval $t$ such that, under the one-shot economic dispatch,
\ben
\item generator $i$ is marginal, \ie $0 < g^{\ED}_{it} < \bar{g}_{it}$, and
\item the ramping constraint  of $(g_{it}^{\ED})$  between interval $t-1$ and $t$ is not binding and that between interval $t$ and $t+1$ binding with all multipliers being positive,
\een
then there does not exist uniform prices $\pibf$ for which $(\Gbf^{\ED}, \pibf)$ is a strong equilibrium.
\label{thm:PE}
\end{theorem}
The  conditions above on the one-shot economic dispatch are mild; they are easily satisfied for stochastic demands over a sufficiently large $T$. As is shown in the appendix of this paper, empirical studies based on a practical network and demand model shows that the above conditions are satisfied by overwhelming majority of cases.

The significance of Theorem~\ref{thm:PE} is that all uniform prices suffer from the lack of dispatch-following incentives. As a result, out-of-the-market uplifts are necessary to ensure that generators follow the operator's dispatch signal.

\subsection{Out-of-the-market settlements}
The {\em out-of-the-market settlement}, also known as {\em uplift}, is a process for the operator to compensate market participants  for inadequate payments due to inaccurate, incomplete, or non-convex  models.  Out-of-the-market settlements are in general discriminative and determined in {\it ex-post} over the entire scheduling horizon $\Hmsc$ \cite{ONeill&etal:05EJOR,Gribik&Hogan&Pope:07,Al-Abdullah&Abdi-Khorsand&Hedman:15TPS}.
Two popular schemes are the make-whole (MW) settlement used in most operators in the U.S. and the lost-of-opportunity-cost (LOC) settlement implemented in ISO-NE.

Let $\pibf$ be the price vector over $\Hmsc$ and $\gbf_i=(g_{i1},\cdots, g_{iT})$ the generation of generator $i$.  The make-whole (MW) payment ${\rm MW}(\pibf,\gbf_i)$ and the lost-of-opportunity cost (LOC) payment $\mbox{LOC}(\pibf,\gbf_i)$ for generator $i$ are defined by, respectively,
\bea \label{eq:uplift}
{\rm MW}(\pibf,\gbf_i) &=& \max\{0, \sum_{t=1}^T (f_{it}(g_{it}) - \pi_t g_{it})\}, \\
{\rm LOC}(\pibf,\gbf_i) &=& Q_i(\pibf) - \sum_{t=1}^T (\pi_t g_{it}-f_{it}(g_{it})),
\eea
where $Q_i(\pibf)$ is the maximum profit the generator would have received if the generator self-schedules for the given price $\pibf$:
\beq \label{eq:Q}
\begin{array}{lll}
Q_i(\pibf)= & \underset{\pbf=(p_1,\cdots, p_T)}{\rm maximize} & \sum_{t=1}^T(\pi_{t}p_t - f_{it}(p_t)) \\[0.5em]
& {\rm subject~to}&  0\leq p_{t} \leq \bar{g}_{i}\\
&  & -\underline{r}_i\leq p_{(t+1)}-p_t \leq \bar{r}_i. \\
\end{array} \hfill
\eeq
It turns out that, when  $Q_i(\pibf) \ge 0$, we always have ${\rm LOC}(\pibf,\gbf_i)  \ge  {\rm MW}(\pibf,\gbf_i)$.  See Appendix D.

%The following property states that, with uplifts, a generator always has nonnegative total surplus and LOC payment is always no smaller than the MW payment.
%\begin{proposition}\label{prop:uplift}
%Let $(\pibf, \gbf)$ be a price-generation pair over the entire scheduling horizon $\Hmsc$ and $S(\pibf,\gbf)$ its in-market surplus, \ie
%\[
%S(\pibf,\gbf) :=  \sum_{t=1}^T (\pi_t g_t  - f_{t}(g_{t})),
%\]
%where $f_{t}(\cdot)$ is the generation cost function in interval $t$.  If $Q(\pi)$ in (\ref{eq:Q}) is non-negative, then
%\beq
%{\rm LOC}(\pibf,\gbf)  \ge  {\rm MW}(\pibf,\gbf) .
%\eeq
%\end{proposition}

The following proposition, an immediate consequence of the general equilibrium conditions, shows that the LOC uplift is a measure of the dispatch-following disincentives.

\begin{proposition}[LOC and general equilibrium]\label{prop:LOC2}
 A dispatch matrix-price pair $(\Gbf=[\gbf_1,\cdots, \gbf_N]^{\intercal},\pibf)$ satisfies the general equilibrium condition if and only if the LOC uplifts for all generators are zero.
\end{proposition}

The following theorem  shows that uniform pricing in general will lead to non-zero LOC.  Therefore, price discrimination is unavoidable in practice.

\begin{theorem}[Uniform pricing and out-of-the-market uplifts] \label{thm:UniformPricingLOC}
Let $\{g_{it}^{\RED}\}$ be the rolling-window economic dispatch over the entire scheduling horizon $\Hc$.  There does not exist a uniform pricing scheme under which all generators have zero LOC if there  exist generators $i$ and  $j$ and interval $t^*  \in \Hc$  such that
\ben
\item generators $i$ and $j$ have different bid-in marginal costs of generation
\[
  \frac{d}{dg} f_{it^*}(g_{it^*}^{\RED})  \ne  \frac{d}{dg} f_{jt^*}(g_{jt^*}^{\RED} );
  \]
\item both generators are ``marginal''  in $t^*$, \ie
\[
 g_{it^*}^{\RED} \in (0,\bar{g}_i),~~g_{jt^*}^{\RED} \in (0, \bar{g}_j);
\]
\item and both generators  have no binding ramping constraints from intervals $t^*-1$ to $t^*$ and from $t^*$ to $t^*+1$.
 \een
\end{theorem}
Note that  the conditions Theorem~\ref{thm:UniformPricingLOC} are stated for the rolling-window dispatch under arbitrary forecast errors.  Note also that condition (2) on the existence of simultaneously marginal generators  can happen because of the rolling-window economic dispatch model.   Empirical evaluations under practical demand models show that conditions (2) and (3) hold in high percentage when the ramping constraints are tight.  See Appendix J.

\section{Temporal Locational Marginal Price} \label{sec:TLMP}
%TLMP:
Because uniform pricing cannot provide dispatch-following incentives in general, we now consider nonuniform pricing mechanisms.  To this end, we extend LMP to the {\em temporal locational marginal price (TLMP)} and establish that TLMP is a strong equilibrium price, thus eliminating out-of-the-market uplifts.

\subsection{TLMP: a generalization of LMP}
We first consider the one-shot TLMP defined over $\Hmsc$; the rolling-window TLMP follows the same way as the rolling-window LMP.

%As in LMP, TLMP prices a load by the marginal cost of satisfying its demand.  Unlike LMP, TLMP prices  the generation from generator $i$ by its contribution to meeting the system load. In particular, we treat generator $i$ as an inelastic negative demand and pay generator $i$ at the marginal benefit of its generation. Roughly speaking,  generator $i$ is paid at the marginal cost to the system when generator $i$ reduces one MW of its generation.

As in LMP, TLMP prices a load by the marginal cost of satisfying its demand.  Unlike LMP, TLMP prices the generation $i$ in interval $t$ by its contribution to meeting system load in interval $t$.  In particular, we treat generator $i$ as a negative demand and pay generator $i$ at the marginal benefit of its generation in interval $t$.  Roughly speaking, generator $i$ is paid at the marginal cost to the system when it reduces 1 MW of generation in interval $t$.

Define a {\em parameterized economic dispatch} by treating $g_{it}$ as a parameter rather than a decision variable in (\ref{eq:ED}).  Let the partial cost be
\[
F_{-it}(\Gbf) := F(\Gbf)- f_{it}(g_{it}),
\]
which excludes  the cost of generator $i$ in interval $t$.  The  parameterized economic dispatch is defined by (\ref{eq:ED}) with $F_{-it}(\Gbf)$ as the cost function and  $\{g_{i't'}, (i',t')\ne (i,t)\}$ as its decision variables.

%Let $\Gbf^*$ be the solution of the parameterized economic dispatch.
%Let $\Gbf^{\ED}$ be the solution of (\ref{eq:ED}).  At $g_{it}=g_{it}^{\ED}$, we have $\Gbf^*=\Gbf^{\ED}$.

\begin{definition}[TLMP]
The TLMP  for the demand in interval $t$ is defined by the marginal cost of meeting the demand:
\[
\pi^{\mbox{\rm\tiny TLMP}}_{0t} := \frac{\partial}{\partial \hat{d}_t} F(\Gbf^{\ED}).
\]
The TLMP for  generator $i$  in interval $t$  is defined by the marginal benefit of generator  $i$ at $g_{it}=g^{\ED}_{it}$:
\[
\pi^{\mbox{\rm\tiny TLMP}}_{it} := -\frac{\partial}{\partial g_{it}} F_{-it}(\Gbf^{\ED}).
\]
\end{definition}

Proposition~\ref{prop:TLMP} gives an explicit expression for TLMP.

\begin{proposition}  \label{prop:TLMP}
Let $\Gbf^{\ED}$  be the solution to the multi-interval economic dispatch in (\ref{eq:ED}) and $(\lambda^*_t,\underline{\mu}_t^*, \bar{\mu}_t^*,\underline{\rho}_t^*, \bar{\rho}_t^*)$ the dual variables associated with  the constraints.  The TLMP for the demand in interval $t$ is given by
 \[
\pi^{\mbox{\rm\tiny TLMP}}_{0t} = \lambda_t^*.
\]
The TLMP for the generator $i$ in interval $t$ is given by
\beq \label{eq:TLMP1}
\pi^{\mbox{\rm\tiny TLMP}}_{it} =  \lambda^*_t+ \Delta_{it}^*,
\eeq
where $\Delta_{it}^* = \Delta \mu_{it}^*-\Delta \mu_{i(t-1)}^* $, and $\Delta \mu_{it}^*:=\bar{\mu}_{it}^*-\underline{\mu}_{it}^*$.
\end{proposition}
The intuition behind the TLMP expression  is evident from a dual perspective of the economic dispatch.  Specifically, the Lagrangian of the one-shot economic dispatch (\ref{eq:ED}) with the optimal multipliers can be written as
\beq \label{eq:Lagrangian}
\Lc = \sum_{i,t}  \bigg(f_{it}(g_{it})- (\lambda_t^* + \Delta_{it}^*)g_{it} + (\bar{\rho}_{it}^*-\underline{\rho}^*_{it}) g_{it}\bigg) + \cdots
\eeq
where the rest of the terms above are independent of $g_{it}$.  It is evident that, with TLMP $\pi^{\TLMP}_{it}:=\lambda_t^*+\Delta_{it}^*$, the multi-interval dispatch decouples into single-interval dispatch problems.   This property has significant ramifications in the equilibrium properties of TLMP.

Proposition~\ref{prop:TLMP} reveals the structure of TLMP as a natural extension of LMP; it adds to the uniform pricing of LMP with a discriminative {\em ramping price}   $\Delta_{it}^*$.  The LMP portion of TLMP is public as it represents the system-wide energy price whereas the private ramping price accounts for the individual ramping capabilities.  Note also that TLMP incurs  no additional computation costs beyond that in LMP.

 Two interpretations of the ramping price $\Delta_{it}^*$  in TLMP are in order.  First, note  that the TLMP expression above is consistent with that in (\ref{eq:TLMP0}); both expressions give the interpretation that the ramping price in TLMP is the increment of  the shadow prices associated with the ramping constraints.

Second,  the ramping price $\Delta_{it}^*$  can be positive or negative. When the ramping price $\Delta^*_{it}>0$, it can be interpreted as an upfront payment for the ramping-induced lost-of-opportunity cost, which ensures that the generator under TLMP is never under-paid below its generation cost.  When it is negative, it has the interpretation of a penalty for the generator's inability to ramp for greater welfare.  See discussions of  Example I in Sec.~\ref{sec:example} and  Proposition 4 and related discussions in Part II \cite{Chen&Guo&Tong:20TPS}.

%
%Third,  as a direct consequence of Theorem~\ref{thm:TLMP1} and Theorem~\ref{thm:R-TLMP},  the ramping price  $\Delta_{it}^*$  can be viewed as the price adjustments around LMP such that, if each generator self schedules, the economic dispatch $\Gbf^{\ED}$ is the profit maximizing schedule.

\subsection{Dispatch-following incentives of TLMP and R-TLMP}
We now consider the equilibrium and dispatch-following incentives. Because TLMP is a nonuniform pricing, the general and partial equilibrium definitions  given in the previous section need to be generalized slightly.
\bitem
\item Instead of having a single price vector for all generator, we now have an individualized price vector $\pibf_i$ for each generator $i$.
\item
The individual rationality conditions extend naturally by replacing $\pi_t$ in (\ref{eq:geq}-\ref{eq:peq}) by $\pi_{it}$.
\eitem

Theorem~\ref{thm:TLMP1} establishes the strong equilibrium property for the one-shot TLMP.

\begin{theorem}[One-shot TLMP as a strong equilibrium price] \label{thm:TLMP1}  When there is no forecasting error, \ie $\hat{\dbf}=\dbf$, the one-shot multi-interval economic dispatch policy $\Gc^{{}^{\tiny\rm ED}}$  and the TLMP policy $\Pc^{{}^{\tiny\rm TLMP}}$ form a strong equilibrium, thus there is no incentive for any generator to deviate from the economic dispatch signal.

In addition, the one-shot TLMP guarantees revenue adequacy for the operator with total merchandising surplus equal to the ramping charge:
 \bea
 \mbox{\rm MS} &:=&\sum_t \pi^{\TLMP}_{0t}d_t-\sum_{i>0,t} \pi^{\TLMP}_{it}g^{\ED}_{it} \nn\\
 &=& \sum_{i,t} (\bar{\mu}^*_{it}\bar{r}_i + \underline{\mu}^*_{it}\underline{r}_i) \ge 0.  \label{eq:MS}
 \eea
\end{theorem}
The intuition behind the above theorem is evident from the Lagrangian of the one-shot economic dispatch (\ref{eq:Lagrangian}).   Because TLMP decouples the temporal dependencies of the multi-interval  dispatch,  the optimal dispatch $g_{it}^*$ should always satisfy the individual rationality condition for all $i$ and $t$.

The non-negative merchandising surplus and (\ref{eq:MS}) are, perhaps, not surprising;  they are  analogous to the same property for LMP when network congestions occur.

What happens when the load forecasts are not accurate? More importantly, is the rolling-window TLMP a strong equilibrium price for the rolling-window dispatch?

\begin{theorem}[R-TLMP as a strong equilibrium price] \label{thm:R-TLMP}
Let $\gbf_i^{\RED}$ be the rolling-window dispatch for generator $i$ and  $\pibf_i^{\RED}$ its rolling-window TLMP.  Then, for all $i$ and under arbitrary demand forecast error,
$(\gbf_i^{\RED},\pibf^{\RTLMP})$ forms a strong equilibrium, and
\beq \label{eq:LOC=0}
\mbox{\rm LOC}(\pibf_i^{\mbox{\tiny R-TLMP}},\gbf^{\mbox{\tiny R-ED}}_i)=0.
\eeq
\end{theorem}
Note that, when a generator has zero  LOC uplift, then the make-whole payment for the generator is also zero. See Appendix D for corresponding proposition.

The above theorem highlights the most significant property of TLMP for practical situations when the load forecasts used in the rolling-window dispatch are not perfect.  There is no uniform pricing policy that can achieve the same.

\subsection{Truthful-bidding Incentives under R-TLMP and R-LMP}
For the single-interval dispatch and pricing problem, it is known that a price-taking generator under LMP  has the incentive to bid truthfully based on its marginal cost of generation.  Here we show that a price-taker's truthful-bidding behavior generalizes to the multi-interval pricing model under R-TLMP, but not under R-LMP.

At the outset, we note that the price-taking assumption is restrictive; it typically applies to an ideal competitive market and rarely holds strictly in practice.  Under LMP, for instance, a generator with the perfect foresight of an oracle can bid in such a way to make itself a marginal generator so that its bid sets the clearing price.  Note also that,  depending on realized demands,  a generator can be a price-taker in some intervals and price-setter in others under LMP or TLMP.  In practice, a generator without market power may reasonably assume that its bid cannot influence the clearing price and derive its bidding strategy {\it ex ante}  based on the price-taking assumption.  It is under such a setting that we consider how a price-taking profit-maximizing generator bid under R-TLMP.

Let $\qbf(\cdot) = (q_{1}(\cdot), \cdots, q_T(\cdot))$ be the true marginal cost of generation  over $T$ intervals of a specific generator\footnote{For brevity,  we drop the generator index.}.  Let $\fbf(\cdot | \thetabf) = (f_t(\cdot |\theta_t))$ be the generator's bid-in cost  (supply) curve  parameterized by  $\thetabf = (\theta_t)$.  Assume that   $\fbf(\cdot | \thetabf^*) = \qbf(\cdot)$.

With demands and bid-in costs  from other generators fixed, let $\gbf^{\RED}(\thetabf)$ be the vector of cleared generation over $T$ intervals by the ISO under R-TLMP.   The profit of the generator is given by
\beq \label{eq:Pi}
\Pi(\thetabf) = (\pibf^{\RTLMP})^{\T} \gbf^{\RED}(\thetabf) - \sum_{t=1}^T q_t(g^{\RED}_t(\thetabf)),
\eeq
where, under the price-taker assumption,  the clearing price $\pibf^{\RTLMP}$  is not a function of $\thetabf$.

The following theorem establishes that $\thetabf=\thetabf^*$  is a  maximum of $\Pi(\thetabf)$ defined in (\ref{eq:Pi}), \ie bidding at the true cost is optimal.

\begin{theorem}[Truthful-bidding incentive of R-TLMP] \label{thm:TLMPBid}
Consider a price-taking generator with convex generation cost $\qbf(\cdot)$.  Under the rolling-window economic dispatch and R-TLMP with arbitrary forcasting error, it is optimal that the generator bids truthfully with its marginal cost of generation.
\end{theorem}

In contrast to R-TLMP, as shown in Appendix L, R-LMP fails to provide truthful-bidding incentives for price-taking generators because  out-of-the-market uplifts are unavoidable under R-LMP and other uniform pricing schemes.  It is  such out-of-the-market uplifts that incentivize strategic behaviors.

%\tcb{Like the case under LMP,  a generator could influence the clearing price under TLMP had the generator known at the time of bidding that it's ramping constraints will be binding in some interval $t_o$.     In practice, however, in the absence of market power and perfect foresight, it is difficult (if not impossible) to bid ahead of the market-clearing to gain profit.  Therefore,  although a generator under TLMP is not strictly a price-taker, it is reasonable for such a generator to derive its {\em ex-ante bidding strategy} under the price-taking assumption.  Under TLMP, a price-taking generator will bid truthfully.}

\section{Illustrative Examples} \label{sec:example}
We consider two examples involving $T=3$ intervals, one for the one-shot dispatch and pricing policies with perfect load forecasts,  the other for the rolling-window policies with inaccurate forecasts.   The toy examples considered in this section are designed to gain insights into the behavior of these pricing mechanisms. The observations drawn from the examples may not hold in general. In all our simulations, we have quantities in MW and prices in \$/MWh, of which the units are dropped hereafter for simplicity. See Part II for more elaborate Monte Carlo simulations \cite{Chen&Guo&Tong:20TPS}.

\begin{table} [h] \label{tab:oneshot}
\caption{\small One-shot economic dispatch, LMP, and TLMP under linear costs. Initial generation  $\gbf[0]={(380, 40)}$. The price for demand $d_t$ is $\pi_t^{\LMP}$.}
\begin{psfrags}
\centerline{
\psfrag{gpp1}[c]{\small $(g_{it}^{\ED},\pi_t^{\LMP},\pi_{it}^{\TLMP})$}
\psfrag{t=1}[c]{\small $t=1$}
\psfrag{t=2}[c]{\small $t=2$}
\psfrag{t=3}[c]{\small $t=3$}
\psfrag{d}[c]{ $d_t$}
\psfrag{g}[c]{\small $\bar{g}_i$}
\psfrag{c}[c]{\small $c_i$}
\psfrag{r}[c]{\small $\underline{r}_i=\bar{r}_i$}
\scalefig{0.5}\epsfbox{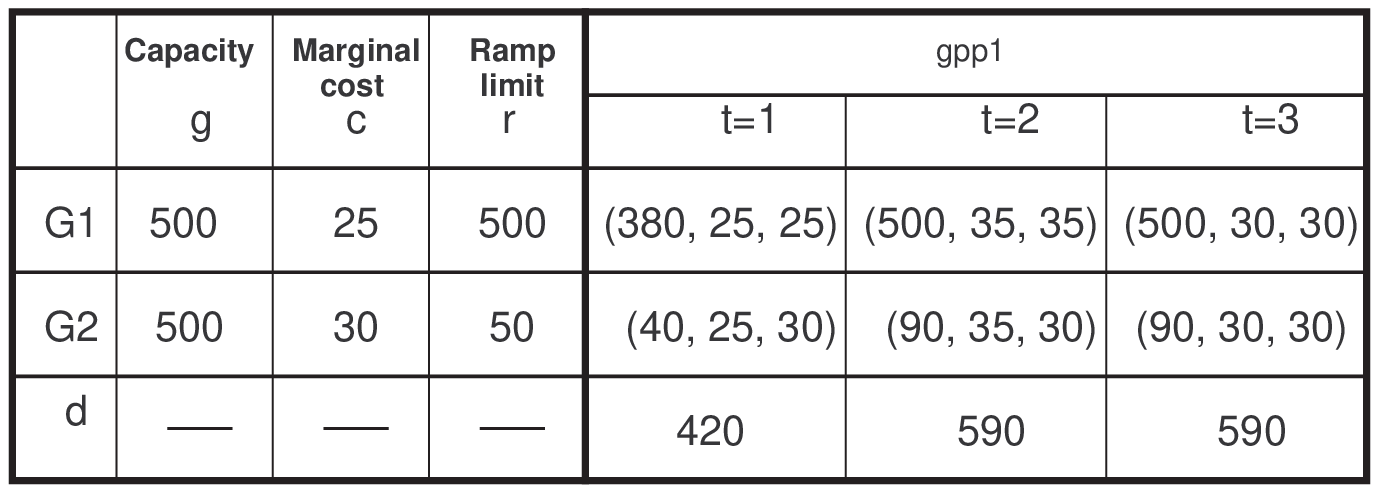}}
\end{psfrags}
\end{table}

\subsection{Example I: one-shot dispatch and pricing}   The economic dispatch, LMP, and TLMP over three intervals are given in the right part of  Table~II.  We make four observations.

First,  G1's ramping limits are not binding over the three intervals.  The LMP and TLMP are the same for G1.

Second, the ramping constraint for G2  is binding between the first and second intervals, making the price of generation under TLMP different from its LMP.
 Note that  in interval $t=1$, G2 is scheduled to generate at the LMP  of \$25/MWh, \$5/MWh   below its marginal cost of \$30/MWh.  As a result, G2 incurs an opportunity cost of \$200  so that it can ramp up to the maximum to the next interval and be paid at \$5/MWh above its marginal cost.  Despite the loss in the first interval, the total surplus over the three intervals  is maximized.  By the general equilibrium property of LMP, there is no incentive for G2 to deviate from the dispatch.

Third, in contrast to LMP, TLMP pays G2 up-front the opportunity cost by adding  \$5/MWh  to the energy price of \$25/MWh.  The up-front payment removes the incentive for G2 to deviate not knowing future demands.   For this reason, the discriminative part of TLMP in (\ref{eq:TLMP1}) has an interpretation as the premium for the ramping-induced opportunity cost.  Note also that, the opportunity cost premium paid to G2 in interval 1 is removed in interval 2.

Fourth, consider the case when the true ramping limit of G2 is 100 MW. Had G2 reported the ramping limit truthfully, G2 would have been dispatched to generate 0 MW in interval 1 and 90 in interval 2 at \$30/MWh  with total profit of zero dollar.  But if G2 falsely declares that it has ramp limit of 50 MW as shown in Table II, we see that G2 under LMP would have made \$250 profit.  This shows that under LMP, there is an incentive for G2 to under-declare its ramp limit. Under TLMP, on the other hand, there is no incentive for G2 to lie about its ramp limit.  See more examples in Part II \cite{Chen&Guo&Tong:20TPS}.

\subsection{Example II: rolling-window dispatch and pricing}
Table~III shows the rolling-window economic dispatch and rolling-window prices with window size $W=2$.  The load forecasts $\hat{\dbf}_t=(\hat{d}_{t}, \hat{d}_{t+1})$ are listed and $\hat{d}_{t}=d_t$ being the actual load.  Note that $\hat{\dbf}_t$ contains forecast errors.

We again make four observations.  First, the missing money scenario happens in this example. G2 is  underpaid by $\pibf^{\RLMP}_1$ in the interval $t=1$.  Unlike the one-shot LMP case, the underpayment is never compensated under R-LMP. The underpayment is compensated out of the market.   The LOC and MW uplifts to G2 are both $\$250$.

{Second, from Table~III, the dispatch of G2 satisfies the conditions in Theorem~\ref{thm:UniformPricingLOC}.  There is no uniform price can remove LOC uplifts.}     For  this example, the argument becomes trivial.  Consider interval $t=1$,  for any price greater than \$25/MWh, G1 self-scheduling  would have generated more than $370$ (MW).  If the price is \$25/MWh, G2 self-scheduling would have generated zero (MW).

Third, for G2 in interval $t=1$, given the inaccurate load forecast of $600$ for interval $t=2$, the rolling-window dispatch for interval $t=2$ is $100$, which makes the ramping constraints from $t=1$ to $t=2$ binding.  The Lagrange multiplier associated with this binding constraint is five.
%Third, for G2 in interval $t=1$, given the inaccurate load forecast of $600$ for interval $t=2$, the rolling-window dispatch for interval $t=2$ is $100$, which makes the ramping constraints from $t=0$ to $t=1$ and from $t=1$ to $t=2$ both binding.  The Lagrange multipliers associated with these two binding constraints are zero and five\footnote{Primal degeneracy occurs in this example. Shadow prices  associated with ramping and power balance constraints are used to compute TLMP.}, respectively.
The  TLMP for  G2 is \$5/MWh above the LMP,   which compensates the underpayment of LMP to the level of marginal cost.  In intervals of $t=2,3$, there are no binding ramping constraints for G2.  G2 is paid at the LMP.  No missing money for TLMP.

Fourth, there is again  no incentive for  G2 to declare its ramp limit untruthfully under TLMP; it will be paid at its marginal costs.
Under LMP, however, there is an incentive for G2 to declare that it has high ramping limits, say 100 MW, and avoid the opportunity cost in the first interval.

\begin{table} [h] \label{tab:rolling}
\caption{\small Rolling-window economic dispatch, LMP, and TLMP. Initial generation $\gbf[0]={(370, 50)}$.  Load is settled at the LMP $\pi_t^{\LMP}$ for all $t$.}
\begin{psfrags}
\centerline{
\psfrag{gpp1}[c]{\small $(g_{it}^{\RED},\pi_t^{\RLMP},\pi_{it}^{\RTLMP})$}
\psfrag{t=1}[c]{\small $t=1$}
\psfrag{t=2}[c]{\small $t=2$}
\psfrag{t=3}[c]{\small $t=3$}
\psfrag{d}[c]{ $\hat{\dbf}_t$}
\psfrag{g}[c]{\small $\bar{g}_i$}
\psfrag{c}[c]{\small $c_i$}
\psfrag{r}[c]{\small $\underline{r}_i=\bar{r}_i$}
\scalefig{0.5}\epsfbox{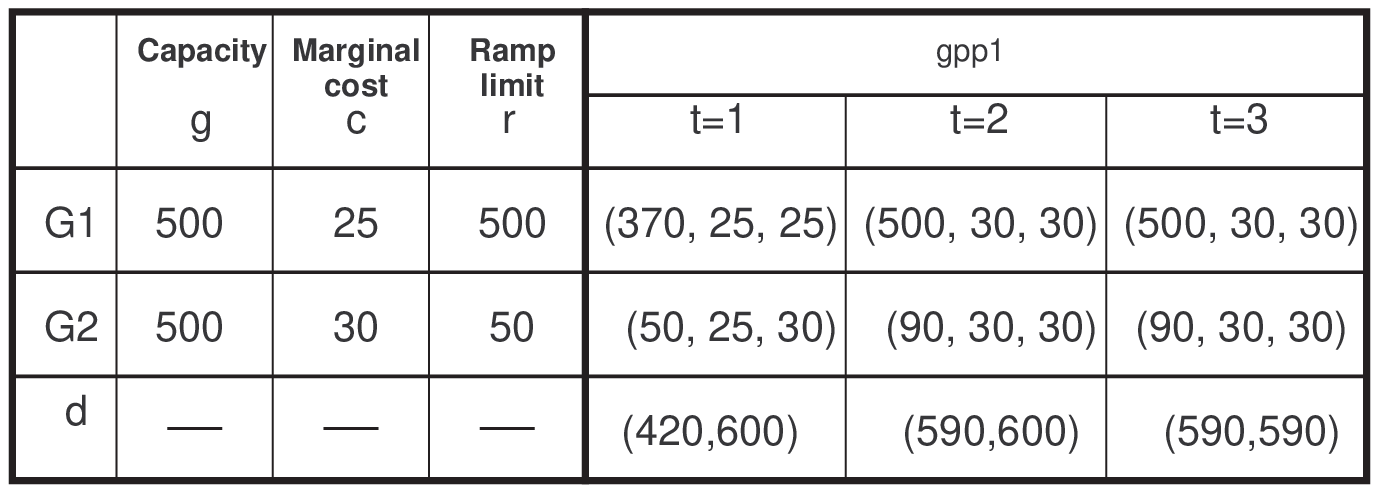}}
\end{psfrags}
%\vspace{-2em}
%\caption{\small Rolling-window economic dispatch, LMP, and TLMP. Initial generation $\gbf[0]={(370, 50)}$.  Load is settled at the LMP $\pi_t^{\LMP}$ for all $t$.}
\end{table}

\vspace{-0.5em}

\section{Discussions}\label{sec:discussion}
We discuss in this section aspects of pricing  multi-interval dispatch  that are not covered in this two-part paper.  The purpose is to provide a broader perspective and contexts beyond the scope of this paper.

We assume a bid-based market model where the market operator collects bids (generation offers) and makes two decisions: one is the allocation of the production levels of the goods (the dispatch over  multiple intervals); the other is setting the prices of generation and consumption.  In analyzing generators' bidding characteristics, we assume that generators are profit-maximizing competitive firms that exhibit price-taking behaviors.  Under such an assumption, we have shown that it is optimal for the generators to bid truthfully under R-TLMP, but not so under R-LMP.

In practice, markets are rarely  competitive, and not all generators are price takers.  To this end, it is more appropriate to model strategic behaviors of generators explicitly.    An excellent example is the work of Hobbs \cite{Hobbs:01TPS} where a Nash-Cournot competition is formulated in analyzing decentralized (bilateral) and centralized (poolco) power markets.  Another example is the work of Philpott, Ferris, and Wets \cite{Philpott&Ferris&Wets:16MP} on the equilibrium, uncertainty, and risk in hydro-thermal systems, which is relevant to the current work for its modeling of inter-temporal constraints and uncertainty.

In  pricing multi-interval economic dispatch with ramping constraints, there is a larger question whether private parameters such as ramping limits, unlike  congestion limits in a public power network, should be modeled explicitly in the operator's pricing decisions.  In this paper, as in some of the recent proposals of ramping products \cite{CAISO_FRP:15,Ela&OMalley:16TPS,Schiro:17FERC,Zhao&Zheng&Litvinov:19arxiv,Hogan:16,Hua&etal:19TPS},  it is the market operator who sets the prices that cover ramping induced costs.   Under LMP and other uniform pricing schemes,  the cost of ramping manifests itself in the form of out-of-the-market uplifts.   For TLMP, on the other hand,  ramping costs show up in the shadow prices of ramping limits within the market clearing process.

An alternative  to the pricing model considered here is to have generators internalize ramping costs in its offer, which is highly nontrivial  \cite{Wilson:02E,Oren:98Bkchap}.  Comparing the two approaches is  outside the scope of this paper.

\section{Conclusion}\label{sec:conclusion}
We have developed a theory for dispatch-following incentives for multi-interval dispatch problems with inter-temporal ramping constraints and
forecast uncertainties.   Since  there is no uniform pricing mechanism that can guarantee dispatch-following incentives without discriminative out-of-the-market uplifts, a non-uniform pricing mechanism such as TLMP can be a valid alternative.  As an extension of LMP, TLMP captures both the energy and the ramping-induced opportunity costs.  As a strong equilibrium pricing mechanism, TLMP guarantees dispatch-following incentives under arbitrary forecast errors and generalizes many properties of LMP.

Evaluating pricing schemes in practice must take into account many factors.  In  Part II of this paper \cite{Chen&Guo&Tong:20TPS}, we conduct more careful simulation studies using relevant performance metrics to compare several benchmark pricing schemes.

\section*{Acknowledgement}
The authors are grateful for the many discussions with Dr. Tongxin Zheng whose insights helped to shape this two-part paper.  We are also benefited from helpful comments and critiques from Shumel Oren, Kory Hedman, Mojdeh Abdi-Khorsand, Timothy Mount, and Bowen Hua.

The authors wish to thank anonymous reviewers and the associate editor  for raising numerous  issues and providing constructive comments, which considerably strengthened this paper during the review process.
{
\bibliographystyle{IEEEtran}
\bibliography{BIB}

\begin{thebibliography}{10}
\providecommand{\url}[1]{#1}
\csname url@samestyle\endcsname
\providecommand{\newblock}{\relax}
\providecommand{\bibinfo}[2]{#2}
\providecommand{\BIBentrySTDinterwordspacing}{\spaceskip=0pt\relax}
\providecommand{\BIBentryALTinterwordstretchfactor}{4}
\providecommand{\BIBentryALTinterwordspacing}{\spaceskip=\fontdimen2\font plus
\BIBentryALTinterwordstretchfactor\fontdimen3\font minus
  \fontdimen4\font\relax}
\providecommand{\BIBforeignlanguage}[2]{{%
\expandafter\ifx\csname l@#1\endcsname\relax
\typeout{** WARNING: IEEEtran.bst: No hyphenation pattern has been}%
\typeout{** loaded for the language `#1'. Using the pattern for}%
\typeout{** the default language instead.}%
\else
\language=\csname l@#1\endcsname
\fi
#2}}
\providecommand{\BIBdecl}{\relax}
\BIBdecl

\bibitem{Guo&Tong:18Allerton}
Y.~{Guo} and L.~{Tong}, ``Pricing multi-period dispatch under uncertainty,'' in
  \emph{2018 56th Annual Allerton Conference on Communication, Control, and
  Computing (Allerton)}, Oct 2018, pp. 341--345.

\bibitem{Tong:19PESGM}
L.~{Tong}, ``Pricing multi-period dispatch under uncertainty,'' in \emph{Proc.
  2019 IEEE PES General Meeting}, August 2019.

\bibitem{Xie&Luo&Obadina:11PESGM}
L.~{Xie}, X.~{Luo}, and O.~{Obadina}, ``Look-ahead dispatch in {ERCOT}: Case
  study,'' in \emph{2011 IEEE Power and Energy Society General Meeting}, July
  2011, pp. 1--3.

\bibitem{Peng&Chatterjee:13FERC}
T.~Peng and D.~Chatterjee, ``Pricing mechanism for time-coupled multi-interval
  real-time dispatch,'' in \emph{FERC Software Conference}, June 2013.

\bibitem{CAISO_FRP:15}
``Flexible ramping product: Revised draft final proposal,'' [ONLINE], available
  (2019/9/9) at
  \url{https://www.caiso.com/Documents/RevisedDraftFinalProposal-FlexibleRampingProduct-2015.pdf},
  December 2015.

\bibitem{Mickey:15}
J.~Mickey, ``Multi-interval real-time market overview,'' [ONLINE], available
  (2019/9/9) at
  \url{http://ercot.com/content/wcm/key_documents_lists/76342/5_Multi_Interval_Real_Time_Market_Overview.pdf},
  October 2015.

\bibitem{Parker:15}
N.~Parker, ``Ramping product design,'' [ONLINE], available (2019/9/9) at
  \url{https://www.spp.org/documents/29342/ramp\%20product\%20design.pdf},
  August 2015.

\bibitem{Schiro:17}
D.~A. Schiro, ``Procurement and pricing of ramping capability,'' [ONLINE],
  available (2019/9/9) at
  \url{https://www.iso-ne.com/static-assets/documents/2017/09/20170920-procurement-pricing-of-ramping-capability.pdf},
  September 2017.

\bibitem{Gribik&Hogan&Pope:07}
P.~R. Gribik, W.~Hogan, and S.~L. Pope, ``Market-clearing electricity prices
  and energy uplift,'' [ONLINE], available (2019/9/9) at
  \url{http://www.lmpmarketdesign.com/papers/Gribik_Hogan_Pope_Price_Uplift_123107.pdf},
  2007.

\bibitem{Zhang&Luh&Litvinov&Zheng&Zhao:09PESGM}
B.~{Zhang}, P.~B. {Luh}, E.~{Litvinov}, {Tongxin Zheng}, and {Feng Zhao}, ``On
  reducing uplift payment in electricity markets,'' in \emph{2009 IEEE/PES
  Power Systems Conference and Exposition}, March 2009, pp. 1--7.

\bibitem{Al-Abdullah&Abdi-Khorsand&Hedman:15TPS}
Y.~M. {Al-Abdullah}, M.~{Abdi-Khorsand}, and K.~W. {Hedman}, ``The role of
  out-of-market corrections in day-ahead scheduling,'' \emph{IEEE Transactions
  on Power Systems}, vol.~30, no.~4, pp. 1937--1946, July 2015.

\bibitem{Zhang&Hedman:19NAPS}
S.~{Zhang} and K.~W. {Hedman}, ``Conditions for ramp rates causing uplift,'' in
  \emph{2019 North American Power Symposium}, October 2019, pp. 1--5.

\bibitem{Wilson:02E}
\BIBentryALTinterwordspacing
R.~Wilson, ``Architecture of power markets,'' \emph{Econometrica}, vol.~70,
  no.~4, pp. 1299--1340, 2002. [Online]. Available:
  \url{http://www.jstor.org/stable/3082000}
\BIBentrySTDinterwordspacing

\bibitem{Ela&OMalley:16TPS}
E.~{Ela} and M.~{O'Malley}, ``Scheduling and pricing for expected ramp
  capability in real-time power markets,'' \emph{IEEE Transactions on Power
  Systems}, vol.~31, no.~3, pp. 1681--1691, May 2016.

\bibitem{Hua&etal:19TPS}
B.~{Hua}, D.~A. {Schiro}, T.~{Zheng}, R.~{Baldick}, and E.~{Litvinov},
  ``Pricing in multi-interval real-time markets,'' \emph{IEEE Transactions on
  Power Systems}, vol.~34, no.~4, pp. 2696--2705, July 2019.

\bibitem{Zhao&Zheng&Litvinov:19arxiv}
J.~Zhao, T.~Zheng, and E.~Litvinov, ``A multi-period market design for markets
  with intertemporal constraints,'' [ONLINE], available (2019/9/9) at
  \url{https://arxiv.org/abs/1812.07034}, June 2019.

\bibitem{Schiro:17FERC}
D.~Schiro, ``Flexibility procurement and reimbursement: a multi-period pricing
  approach,'' in \emph{FERC Software Conference}, June 2017.

\bibitem{Hogan:16}
W.~Hogan, ``Electricity market design: Optimmization and market equilibrium,''
  [ONLINE], available (2019/9/9) at
  \url{https://sites.hks.harvard.edu/fs/whogan/Hogan_UCLA_011316.pdf}, January
  2016.

\bibitem{Chen&Guo&Tong:20TPS}
C.~{Chen}, Y.~{Guo}, and L.~{Tong}, ``Pricing multi-interval dispatch under
  uncertainty part {II}: Generalization and performance,'' \emph{IEEE
  Transactions on Power Systems}, 2020 (early access), see an extended version
  at \url{https://arxiv.org/abs/1912.13469}.

\bibitem{Mas-Colell&Winston&Green:95book}
A.~Mas-Colell, M.~D. Whinston, and J.~R. Green, \emph{Microeconomic
  Theory}.\hskip 1em plus 0.5em minus 0.4em\relax New York: Oxford University
  Press, 1995.

\bibitem{Wu&Variaya&Spiller&Oren:96JRE}
F.~Wu, P.~Varaiya, P.~Spiller, and S.~Oren, ``Folk theorems on transmission
  access: Proofs and counter examples,'' \emph{Journal of Regulatory
  Economics}, vol.~10, no.~1, pp. 5--23, July 1996.

\bibitem{ONeill&etal:05EJOR}
\BIBentryALTinterwordspacing
R.~P. O'Neill, P.~M. Sotkiewicz, B.~F. Hobbs, M.~H. Rothkopf, and W.~R.
  Stewart, ``Efficient market-clearing prices in markets with nonconvexities,''
  \emph{European Journal of Operational Research}, vol. 164, no.~1, pp. 269 --
  285, 2005. [Online]. Available:
  \url{http://www.sciencedirect.com/science/article/pii/S0377221703009196}
\BIBentrySTDinterwordspacing

\bibitem{Hobbs:01TPS}
B.~E. {Hobbs}, ``Linear complementarity models of nash-cournot competition in
  bilateral and poolco power markets,'' \emph{IEEE Transactions on Power
  Systems}, vol.~16, no.~2, pp. 194--202, May 2001.

\bibitem{Philpott&Ferris&Wets:16MP}
A.~Philpott, M.~Ferris, and R.~Wets, ``Equilibrium, uncertainty and risk in
  hydro-thermal electricity systems,'' \emph{Mathematical Programming}, vol.
  157, 01 2016.

\bibitem{Oren:98Bkchap}
S.~S. Oren, ``Authority and responsibility of the {ISO}: Objectives, options
  and tradeoffs,'' in \emph{Designing competitive electricity markets},
  H.~po~Chao and H.~G. Huntington, Eds.\hskip 1em plus 0.5em minus 0.4em\relax
  New York: Springer Scince and Business Media, LLC, 1998, ch.~5, pp. 79--96.

\end{thebibliography}
}

\vspace{1em}
\section*{Appendix}
\subsection{Preliminaries}
We derive a more compact vector-matrix representation of LMP, TLMP and associated representations.
For convenience, we focus on scheduling window $\Hmsc=\{1,\cdots, W\}$.  Let the demand (or forecasted demand) be $\dbf=(d_1,\cdots,d_W)$ be the demand in $\Hmsc$, $\gbf_i=(g_{i1}, \cdots, g_{iW})$ the generation of generator $i$, and $\Gbf^{\intercal}=[\gbf_1,\cdots,\gbf_N]$ the generation matrix.
The $W$-interval economic dispatch in the vector-matrix form is defined by
\beq \label{eq:ED1}
\begin{array}{lrl}
\Gc^{{}^{\tiny\rm ED}}: &\underset{\{\Gbf\}}{\rm minimize}   &  F(\Gbf)=\sum_i f_i(\gbf_i)  \\
&  \mbox{subject to} & \mbox{for all $1 \le i \le N$} \\
& \lambdabf : & \Gbf^{\intercal} {\bf 1} = \dbf \\
& (\underline{\rhobf}_{i},\bar{\rhobf}_i):   & {\bf 0} \le \gbf_i \le \bar{\gbf}_{i}, \\
& (\underline{\mubf}_{i},\bar{\mubf}_{i}):  & -\underline{\rbf}_i\le \Abf \gbf_i \le \bar{\rbf}_{i}, \\
\end{array}
\eeq
where $f_i(\gbf_i)=\sum_t f_{it}(g_{it})$ is the total cost for generator $i$,
$\lambdabf=(\lambda_1, \cdots, \lambda_W)$, the vector of dual variables for the equality constraints and $(\underline{\rhobf}_{i}, \bar{\rhobf}_{i},\underline{\mubf}_{i},\bar{\mubf}_{i})$ vectors of dual variables for the inequalities associated generator $i$, and $\Abf$  is a $W\times W$  {\em lower bi-digonal matrix} with 1 on the diagonals and -1 on the off diagonals.

Let the Lagrangian of $\Gc^{{}^{\tiny\rm ED}}$ be
\bea
L &=& \sum_i f_i(\gbf_i) + \lambdabf^{\intercal} (\dbf-\Gbf^{\intercal}{\bf 1})\nn\\
& & + \sum_i \bigg(\bar{\mubf}_i^{\intercal}(\Abf \gbf_i-\bar{\rbf}_i)-\underline{\mubf}_i^{\intercal}(\Abf \gbf_i+\underline{\rbf}_i)\bigg)\nn\\
& & + \sum_i \bigg(\bar{\rhobf}_i^{\intercal}(\gbf_i-\bar{\gbf}_i)-\underline{\rhobf}_i^{\intercal} \gbf_i\bigg).
\eea

Let $(\Gbf^{\ED}, \lambdabf^*, \underline{\rhobf}^*_{i}, \bar{\rhobf}^*_{i},\underline{\mubf}^*_{i},\bar{\mubf}^*_{i})$ be the solution to $\Gc^{{}^{\tiny\rm ED}}$.
The KKT condition gives
\bea \label{eq:KKT}
\nabla f_i(\gbf_i^*) - \lambdabf^* + \Abf^{\intercal}\Delta\mubf_i^* + \Delta\rhobf_i^* = {\bf 0},
\eea
where $\Delta\mubf_i^*=\bar{\mubf}^*_i - \underline{\mubf}^*_i$ and $\Delta\rhobf_i^*=\bar{\rhobf}^*_i - \underline{\rhobf}^*_i$.

The vector form of the multi-interval LMP and TLMP of generator $i$  are given by, respectively,
\beq\label{eq:LMPTLMP}
\pibf^{\LMP} = \lambdabf^*,~~\pibf^{\TLMP}_i=\lambdabf^*-\Abf^{\intercal}\Delta\mubf^*_i.
\eeq

For the individual rationality condition, for generator $i$, we have the following profit maximization problem for given price $\pibf$:
 \beq \label{eq:Gtilde}
\begin{array}{rll}
\tilde{\Gc}_i:&  \underset{\gbf}{\rm minimize} &  f_i(\gbf)-\gbf^{\intercal} \pibf  \\[0.25em]
 & {\rm subject~to}&
 (\underline{\etabf},\bar{\etabf}):~~ -\underline{\rbf}_i\leq \Abf\gbf \leq \bar{\rbf}_i, \\
 & & (\underline{\zetabf},\bar{\zetabf}):~~   {\bf 0}\leq \gbf \leq\bar{\gbf}_{i}.
\end{array} \hfill
\eeq

By the KKT condition, the solution to the above must satisfy
\beq \label{eq:IRcondition}
\nabla f_i(\gbf)-\pibf +\Abf^{\intercal}\Delta\etabf + \Delta\zetabf ={\bf 0},
\eeq
where $\Delta\etabf=\bar{\etabf} - \underline{\etabf}$ and $\Delta\zetabf=\bar{\zetabf} - \underline{\zetabf}$.

\subsection{Proof of Theorem~\ref{thm:LMP}} %Theorem 1
Let $\Gbf^{\ED}$ be the one-shot economic dispatch and $\pibf^{\LMP}$  the  LMP.  The market clearing condition is already satisfied by $\Gbf^{\ED}$.   The individual rationality condition (\ref{eq:IRcondition}) holds by setting
$(\gbf=\gbf^{\ED}_i, \Delta\etabf_i=\Delta\mubf_i^*, \Delta\zetabf_i=\Delta\rhobf_i^*)$.   \hfil\QED

%Theorem 2
%First, we show that if $(\Gbf,\pibf)$ is a strong equilibrium, then $\Gbf=\Gbf^{\ED}$ and $\pibf=\pibf^{\LMP}$.  Next we show that, under the assumptions on $\Gbf^{\ED}$, $(\Gbf^{\ED},\pibf^{\LMP})$ cannot be a strong equilibrium.
%
%Consider a fixed interval $t$, let $\gbf[t]=(g_{1t},\cdots, g_{Nt})$ be the generation vector at time $t$.  The strong equilibrium condition of $(\Gbf,\pi)$ implies that $g_{it}$ satisfies the solution of individual rationality condition
%\[
%\min_{g\in \Xmsc_{it}}  f_{it}(g)-\pi_t g~~\mbox{for all $i,t$,}
%\]
%where $\Xmsc_{it}$ generation constraints for generator $i$ and interval $t$.  This means that  $\gbf[t]$ is the solution of
% \[
%\min_{p_{it} \in \Xmsc_{it}}  \sum_i f_{it}(p_{it})~~\mbox{subject to $~~\sum_i p_{it}=d_t$}
%\]
%for all $t$ since $\gbf[t]$ clears the market in interval $t$.   This implies that $\Gbf$ is the solution of the one-shot economic dispatch, and $\Gbf=\Gbf^{\ED}$
% by the uniqueness assumption.

\subsection{Proof of Theorem~\ref{thm:UniformPricingLOC}}
Let $\pibf = (\pi_t)$ be an arbitrary uniform price and $\gbf_i^{\RED}$ the rolling-window economic dispatch of generator $i$.
 If generator $i$ has zero LOC under $\pibf$, then $\gbf_i^{\RED}$ must satisfy
the KKT conditions of its LOC optimization:
\[
\nabla f_i(\gbf^{\RED}_i) = \pibf- \Abf^{\top} \Delta \etabf_i - \Delta \zetabf_i,
\]
where $\Delta \etabf_i $ and $\Delta \zetabf_i$ are Lagrange multipliers associated with the LOC optimization.

If condition (2) and (3)  of Theorem~\ref{thm:UniformPricingLOC}  are satisfied for generator $i$  in interval $t^*$, the respective multipliers  in $\Delta\etabf_i$ associated  ramping at $t^*$ and generation limits $\Delta\zetabf_i$  must be zero, which implies
      \[
    \frac{d}{dg} f_{it^*}(g_{it^*}^{\RED})  = \pi_{t^*}.
    \]

Likewise, if generator $j\ne i$ also satisfies the same two conditions in the same interval $t^*$, we must have
\[
 \frac{d}{dg} f_{it^*}(g_{it^*}^{\RED})  =  \frac{d}{dg} f_{jt^*}(g_{jt^*}^{\RED})  = \pi_{t^*},
 \]
which contradicts to the fact that the  two generators have different marginal bid-in costs of generation in interval $t^*$.
\hfil\QED

\subsection{LOC vs. MW uplifts} \label{sec:LOCMW}

\begin{proposition}\label{prop:uplift}
Let $(\pibf, \gbf)$ be a price-generation pair over the entire scheduling horizon $\Hmsc$ and $S(\pibf,\gbf)$ its in-market surplus, \ie
\[
S(\pibf,\gbf) :=  \sum_{t=1}^T (\pi_t g_t  - f_{t}(g_{t})),
\]
where $f_{t}(\cdot)$ is the generation cost function in interval $t$.  If $Q(\pi)$ in (\ref{eq:Q}) is non-negative, then
\beq
{\rm LOC}(\pibf,\gbf)  \ge  {\rm MW}(\pibf,\gbf) .
\eeq
\end{proposition}

{\em Proof:}  By the definition of ${\rm MW}(\pibf,\gbf)$,
\[
{\rm MW}(\pibf,\gbf)  + S(\pibf,\gbf) = \max\{0,S(\pibf,\gbf)\} \ge 0.
\]
By the definition of ${\rm LOC}(\pibf,\gbf)$ with $Q(\pibf)$ is defined in (\ref{eq:Q}),
\bea
{\rm LOC}(\pibf,\gbf)+S(\pibf,\gbf)&=&Q(\pibf) \nn\\
&\ge& \max\{0,S(\pibf,\gbf)\} \nn\\
&=& {\rm MW}(\pibf,\gbf)  + S(\pibf,\gbf).\nn
\eea
 \QED

\subsection{Proof of Proposition~\ref{prop:TLMP}}
TLMP for demand $\hat{d}_t$ is same as LMP; it is  defined by the marginal cost of serving $\hat{d}_t$:
\[
\pi_{0t}^{\TLMP} := \frac{\partial}{\partial \hat{d}_t} F(\Gbf^{\ED}) = \lambda_t^*.
\]

To compute TLMP for generator $i$ in interval $t$, consider the modified multi-interval economic dispatch with generator $i$ in interval $t$ fixed at the optimal economic dispatch level, $g_{it}=g_{it}^{\ED}$:
\beq \label{eq:EDa}
\begin{array}{lrl}
\Gc': &\underset{\{\Gbf=[g_{jt'}, (j,t') \ne (i,t) ]\}}{\rm minimize}   &  F_{-it}(\Gbf)  \\
&  \mbox{subject to} & \mbox{for all $j$ and $t' \in \Hmsc $}\\
& \lambda_{it'}: & \sum \limits_{j}^N g_{jt'}= d_{t'},\forall t' \in \Hmsc \smallsetminus \{t\}  \\
%\mbox{for all $i$ and $t=t_o:t_o+W-1$}\\
& (\underline{\gamma}_{jt'},\bar{\gamma}_{jt'}):   & 0 \le g_{jt'} \le \bar{g}_{j},\forall (j,t') \ne (i,t), \\
%& & \hfill \forall (j,t') \ne (i,t),\\
& (\underline{\eta}_{jt'},\bar{\eta}_{jt'}):  &  -\underline{r}_j\le g_{j(t'+1)}-g_{jt'} \le \bar{r}_{j},\\
& & \hfill \forall j\ne i,\\
& (\underline{\eta}_{it'},\bar{\eta}_{it'}):  &  -\underline{r}_i\le g_{i(t'+1)}-g_{it'} \le \bar{r}_{i}, \\
& & \hfill \forall t' \in \Hmsc \smallsetminus \{t-1,t\},\\
& \lambda_{it}: & \sum \limits_{j\ne i}^N g_{jt}= d_{t} - g_{it}^{\ED},  \\
%& (\underline{\gamma}_{jt},\bar{\gamma}_{jt}):   & 0 \le g_{jt} \le \bar{g}_{j}, \forall j\ne i \\
& (\underline{\eta}_{it},\bar{\eta}_{it}):  &  -\underline{r}_i\le g_{i(t+1)}-g^{\ED}_{it} \le \bar{r}_{i}, \\
& (\underline{\eta}_{i(t-1)},\bar{\eta}_{i(t-1)}):  &  -\underline{r}_i\le g^{\ED}_{it}-g_{i(t-1)} \le \bar{r}_{i}. \\
\end{array}
\eeq

 By the envelope theorem, at the optimal solution $\Gbf^*=[g_{it}^*]$ and
 $(\underline{\gamma}_{it}^*,\bar{\gamma'}_{it}^*,\underline{\eta'}_{it}^*,\bar{\eta'}_{it}^*)$
 of $\Gc_{t_o}'$, we have
\bea
-\frac{\partial}{\partial g_{it}^*} F_{-it}(\Gbf^*) &=& \lambda^*_{it} + \Delta \eta_{it}^* - \Delta \eta_{i(t-1)}^*\nn\\
&=& \lambda_t^* + \Delta_{it}^*, \nn
\eea
where, for the last equality, {we have $\lambda_{it}^* = \lambda_t^*, \eta_{it}^*=\mu_{it}^*$ at the optimal dispatch defined in (\ref{eq:ED}).} \hfil\QED

\subsection{Proof of Theorem~\ref{thm:TLMP1}} We first show that $(\Gbf^{\ED},(\pibf_i^{\mbox{\tiny TLMP}}))$ satisfies the general equilibrium conditions. Again, we only need to check the individual rationality condition since the economic dispatch $\Gbf^{\ED}$ already satisfies the market clearing condition as well as all the ramping constraints.

For the individual rationality condition, we consider the optimization $\tilde{\Gc}_i$ (\ref{eq:Gtilde}) with $\pibf=\pibf^{\mbox{\tiny TLMP}}$.  Setting $\underline{\etabf}=\bar{\etabf}={\bf 0}$ and $\Delta{\zetabf}=\Delta \rhobf^*_i$, by the KKT condition, $\gbf_i^{\ED}$ is a solution to $\tilde{\Gc}_i$.  Thus $(\pibf_i^{\mbox{\tiny TLMP}},\gbf_i^{\ED})$ satisfies the individual rationality condition for all $i$. 

To show that $(\Gbf^{\ED},(\pibf_i^{\mbox{\tiny TLMP}}))$ also satisfies the strong equilibrium condition, we note that $(\Gbf^{\ED}, \bar{\etabf}_i=\underline{\etabf}_i={\bf 0}, \bar{\rhobf}_i^*, \underline{\rhobf}_i^*)$ is a solution to (\ref{eq:Gtilde}).  Because the dual variables for ramping constraints are all zero, the multi-interval optimization decouples in time under $\pibf_i^{\mbox{\tiny TLMP}}$. We have $q_{it}^{\RED}$ as a solution to (\ref{eq:peq}) for individual rationality.

To show the revenue adequacy for the operator, we compute the merchandising surplus under TLMP.  From (\ref{eq:LMPTLMP}),
\bea
\mbox{MS} &=& \dbf^{\intercal} \lambdabf^{\LMP} - \sum_i (\lambdabf^{\LMP}-\Abf^{\intercal}\Delta\mubf_i^*)^{\intercal} \gbf_i^{\ED}\nn\\
&=&  \sum_i (\Delta\mubf_i^*)^{\intercal} \Abf \gbf_i^{\ED} \nn\\
&=& \sum_{i} \bar{\rbf}_i^{\intercal} \bar{\mubf}^*_i + \underline{\rbf}_i^{\intercal} \underline{\mubf}_i^* \ge 0,\nn
\eea
where the last equality comes from the complementary slackness condition. \hfill\QED

\subsection{Proof of Theorem~\ref{thm:R-TLMP}}
Within this proof, we will focus on a particular generator, say generator $i$. For brevity, we drop the subscript $i$ of all variables associated with generator $i$.

Let $\gbf^{\RED}=(g_1^{\RED},\cdots,g_T^{\RED})$ be the rolling-window economic dispatch over $\Hmsc$ and $\pibf^{\RTLMP}=(\pi_1^{\RTLMP},\cdots,\pi_T^{\RTLMP})$  the rolling-window TLMP vector.

%At $t=1$, let $\dbf_1=(d_{11},\cdots, d_{1W})$ be the load forecast with $d_{11}=d_1$ being the actual demand in interval $t=1$ and $d_{1k}, k>1$ the demand forecasts.  In general, $d_{1k} \ne d_k$.

Let $\gbf_t^{\mbox{\tiny ED}}$ be the $W$-window economic dispatch at time $t$ over $\Hmsc_t$ from (\ref{eq:ED1}) based on
$\dbf_t=(d_{t1},\cdots,d_{tW})$.  Note that $d_{t1}=d_t$, the actual demand for interval $t$, and the rest of entries of $\dbf_t$ are forecasts with errors.  Let $\pibf_t^{\mbox{\tiny TLMP}}$ be the corresponding TLMP vector given in (\ref{eq:TLMP1}).

From the proof of Theorem~\ref{thm:TLMP1} (with $T=W$), the profit maximization,
 \beq
\begin{array}{rll}
\tilde{\Gc}_t:&  \underset{\gbf=(g_1,\cdots,g_W)}{\rm minimize} &  (f_t(\gbf)-\gbf^{\intercal} \pibf_t^{\TLMP})  \\[0.25em]
 & {\rm subject~to}&
 (\underline{\etabf},\bar{\etabf}):~~ -\underline{\rbf}_t\leq \Abf\gbf \leq \bar{\rbf}_t, \\
 & & (\underline{\zetabf},\bar{\zetabf}):~~   {\bf 0}\leq \gbf \leq\bar{\gbf}_{t},
\end{array} \hfill
\eeq
has a solution $\gbf^{\ED}_t$ with $\underline{\etabf}=\bar{\etabf}={\bf 0}$, where
$f_t(\gbf)$ is the generation cost over $\Hmsc_t$.   This means that $\gbf_t^{\ED}$ is a solution to the ramp-unconstrained optimization
\[
\gbf_t^{\ED}=\arg\min_{{\bf 0}\le \gbf \le \bar{\gbf}_t}  (f(\gbf)-\gbf^{\intercal}\pibf_t^{\TLMP}).
\]

By the rolling-window dispatch and pricing policies, the first entry  of $\gbf_t^{\mbox{\tiny ED}}$ is  $g_{t}^{\mbox{\tiny R-ED}}$---the dispatch that is implemented in interval $t$---and the first entry  of $\pibf_t^{\mbox{\tiny TLMP}}$ is the the rolling-window price $\pi_{t}^{\mbox{\tiny R-TLMP}}$ in interval $t$.  We thus have
\beq \label{eq:gtRED}
g_{t}^{\RED}=\arg\min_{0\le g \le \bar{g}}( f_t(g)-g\pi_{t}^{\RTLMP}),
\eeq
which implies that $\gbf^{\RED}$ is the solution to the ramp-unconstrained optimization
\[
\gbf^{\RED} = \arg\min_{{\bf 0}\le  \gbf \le \bar{\gbf}} ( f(\gbf)-\gbf^{\intercal}\pibf^{\RTLMP}).
\]

Let $\gbf^*$ be the solution to the (ramp-constrained) LOC optimization (\ref{eq:Gtilde}) with $\pibf=\pibf^{\RTLMP}$, we must have
\[
f(\gbf^{\RED})-(\gbf^{\RED})^{\intercal}\pibf^{\RTLMP} \le f(\gbf^*)-(\gbf^*)^{\intercal}\pibf^{\RTLMP}.
\]

Note, however, that $\gbf^{\RED}$ satisfies all the constraints in (\ref{eq:Gtilde}), the above inequality holds with equality, and  $\gbf^{\RED}$ is a solution to (\ref{eq:Gtilde}). Therefore, $\mbox{LOC}(\gbf^{\RED},\pi^{\RTLMP})=0.$

By Proposition~\ref{prop:LOC2}, $(\Gbf^{\RED}, \Pibf^{\RTLMP})$ is a general equilibrium. From (\ref{eq:gtRED}), we conclude that $(\Gbf^{\RED}, \Pibf^{\RTLMP})$ also satisfies  the strong equilibrium conditions. \hfill \QED

\subsection{{Proof of Theorem~\ref{thm:TLMPBid}}}  We focus on a specific generator, henceforth dropping the generator index in the notation within this proof.   Under the price-taker assumption, from (\ref{eq:Pi}), we have
\[
\Pi(\thetabf^*) = (\pibf^{\RTLMP})^{\T} \gbf^{\RED}(\thetabf^*) - \sum_{t=1}^T q_t(g^{\RED}_t(\thetabf^*)).
\]
From Theorem~\ref{thm:R-TLMP}, we know that, when bidding truthfully, there will be no LOC, which implies that
\[
\Pi(\thetabf^*) \ge (\pibf^{\RTLMP})^{\T} \gbf - \sum_{t=1}^T q_t(\gbf),
\]
 for every $\gbf$ in the profit maximization problem.  Because a price-taker's bid can only influence dispatch $\gbf^{\RED}(\theta)$, we have $\Pi(\thetabf^*) \ge \Pi(\thetabf)$. \hfil\QED

\begin{figure}[h]
\center
\begin{psfrags}
\scalefig{0.5}\epsfbox{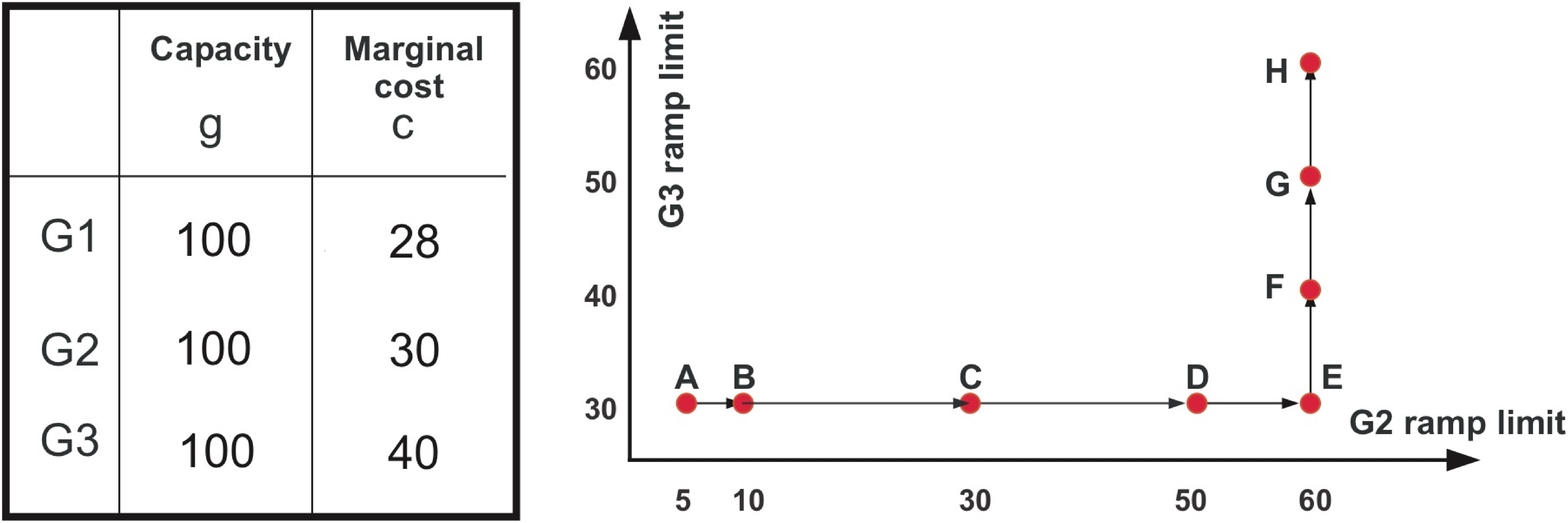}
\scalefig{0.24}\epsfbox{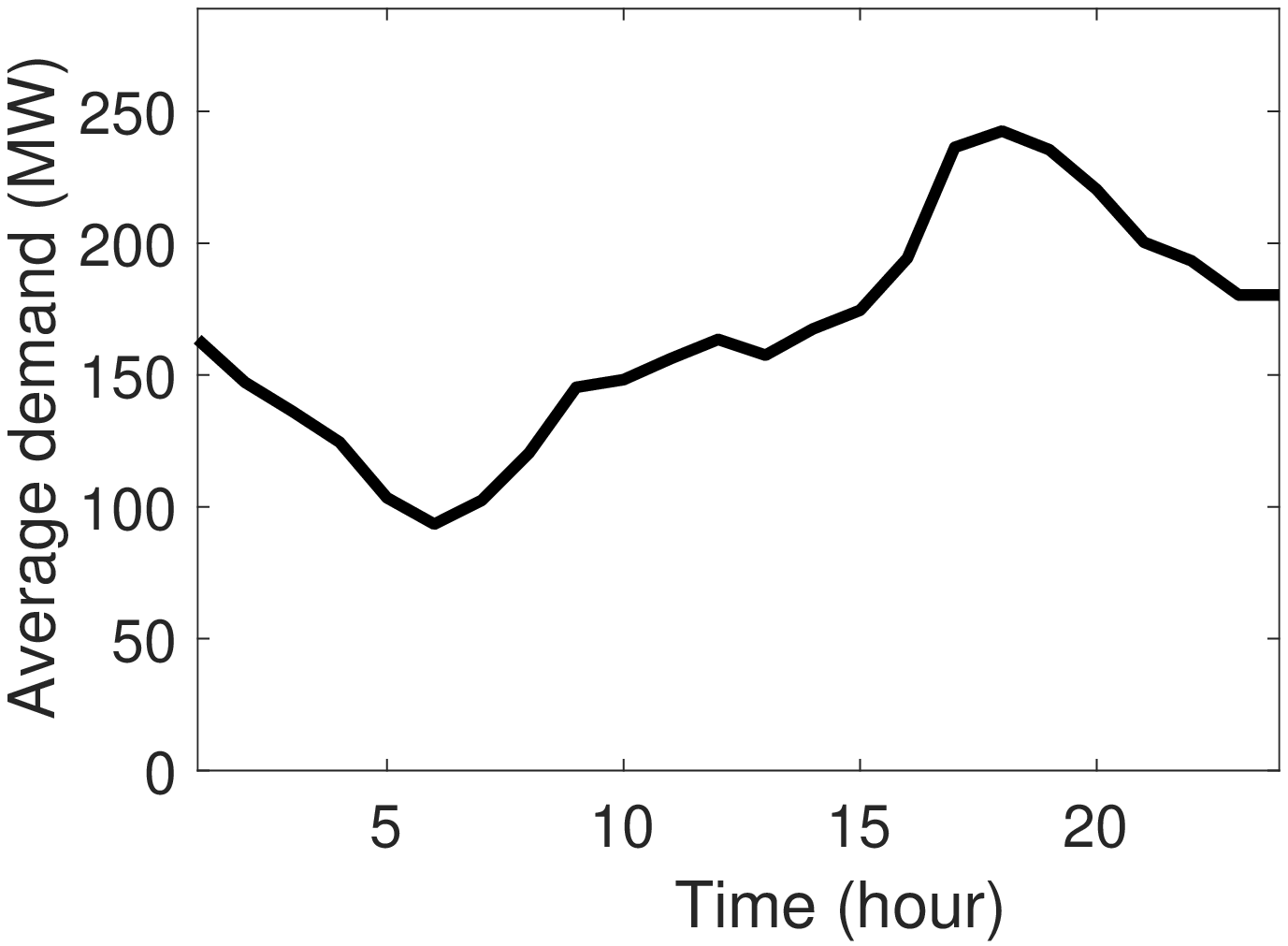}\scalefig{0.24}\epsfbox{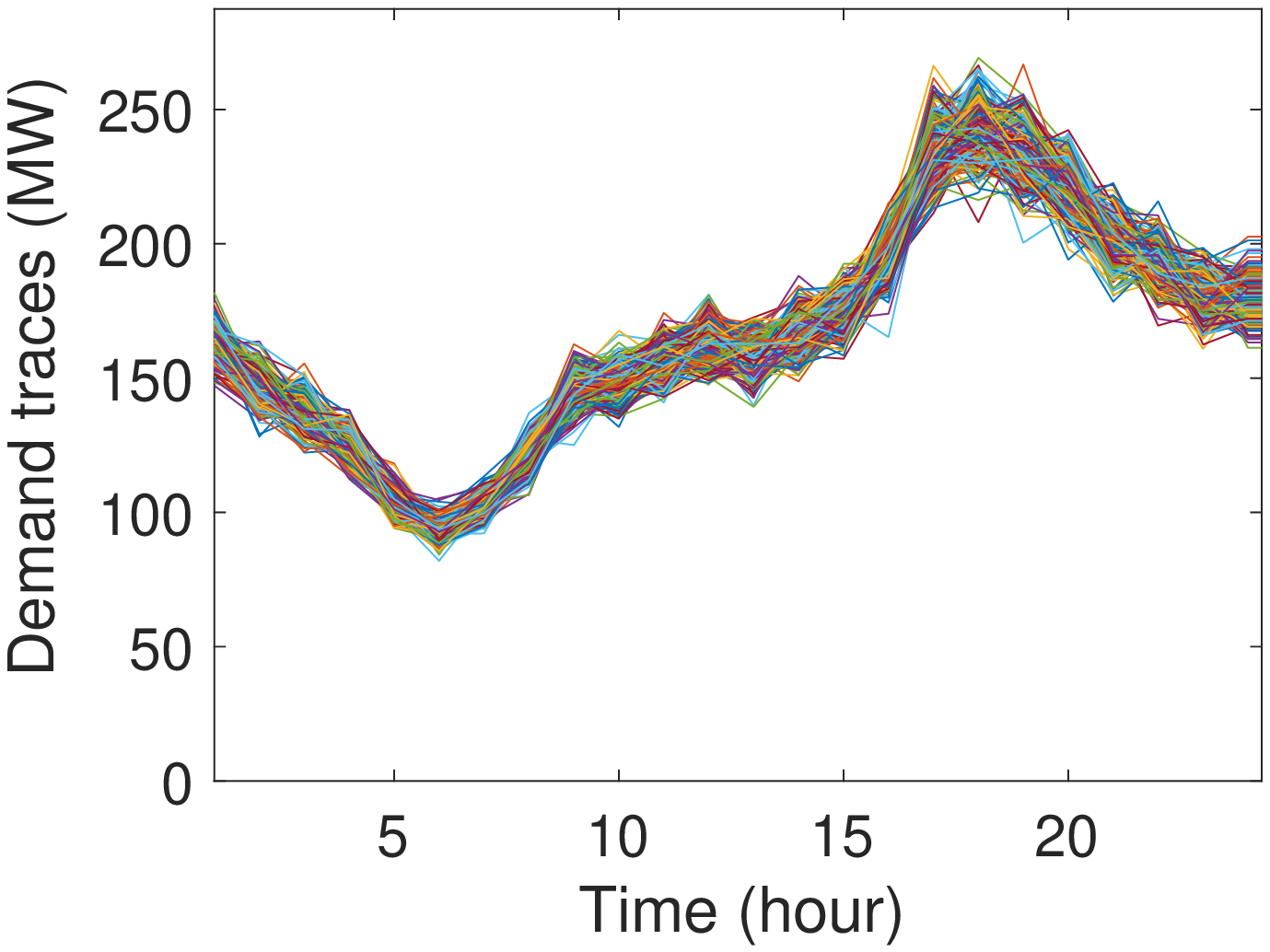}
\end{psfrags}
\vspace{-1em}\caption{\small Top left: generator parameters.  The ramp limit for G1 is fixed at 25 (MW/h).  Top right: a path of ramping events. Bottom left: average demand.  Bottom right: demand traces.}
\label{fig:casesetting}
\end{figure}

\subsection{Simulations on the conditions in Theorem~\ref{thm:UniformPricingLOC}}\label{sec:THM2AssumTest}
We present empirical test results on how  frequently assumptions  in Theorem~\ref{thm:UniformPricingLOC}  of Part I hold.  Fig.~\ref{fig:casesetting} shows the parameters of the generators and load scenarios in this three-generator-single-bus case. We evaluated assumptions under different ramping limits along the path from scenarios A to H, where scenarios A had the most stringent ramping constraints and H the most relaxed.  Moreover, we evaluated assumptions under different load forecast errors with a standard forecasting error model\footnote{The forecast $\hat{d}_{(t+k)|t}$ at $t$  of demand $d_{t+k}$ is $\hat{d}_{(t+k)|t}=d_{t+k}+\sum_{i=1}^k \epsilon_k$ where $\epsilon_k$ is i.i.d. Gaussian with zero mean and variance $\sigma^2$.}, where the demand forecast $\hat{d}_{(t+k)|t}$ of $d_{t+k}$  at time $t$ had error variance $k\sigma^2$  increasing linearly with $k$. And $\sigma$ varied from $\sigma=0\%$ to $\sigma=6\%$.This simulation setting was the same with cases in \cite{Chen&Guo&Tong:20TPS}, and 400 realizations with a standard deviation of 4\% were tested with rolling-window optimization over the 24-hour scheduling period, represented by 24 time intervals. And the window size is four intervals in each rolling window optimization.

\begin{figure}[h]
\center
\begin{psfrags}
\scalefig{0.25}\epsfbox{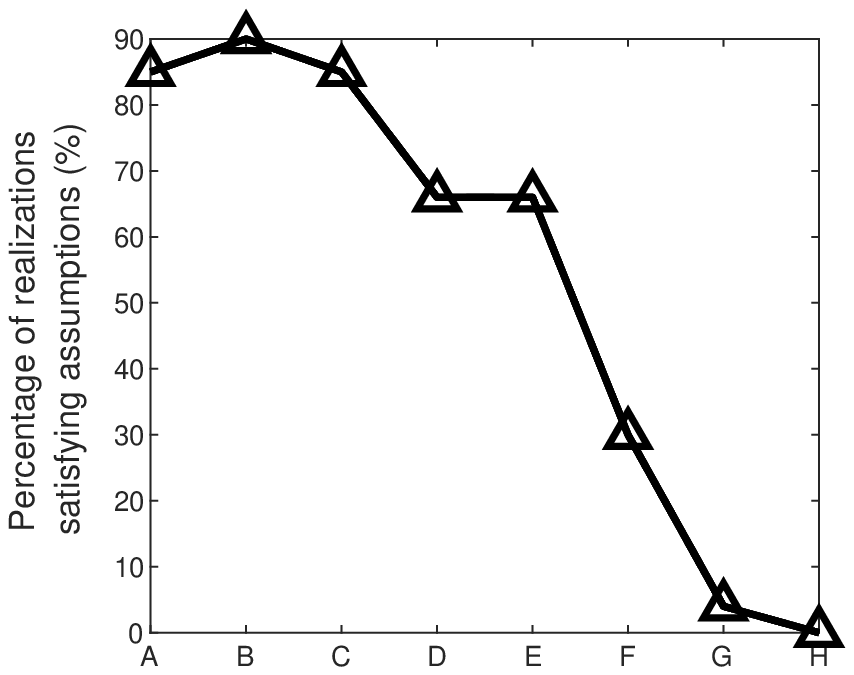}\scalefig{0.25}\epsfbox{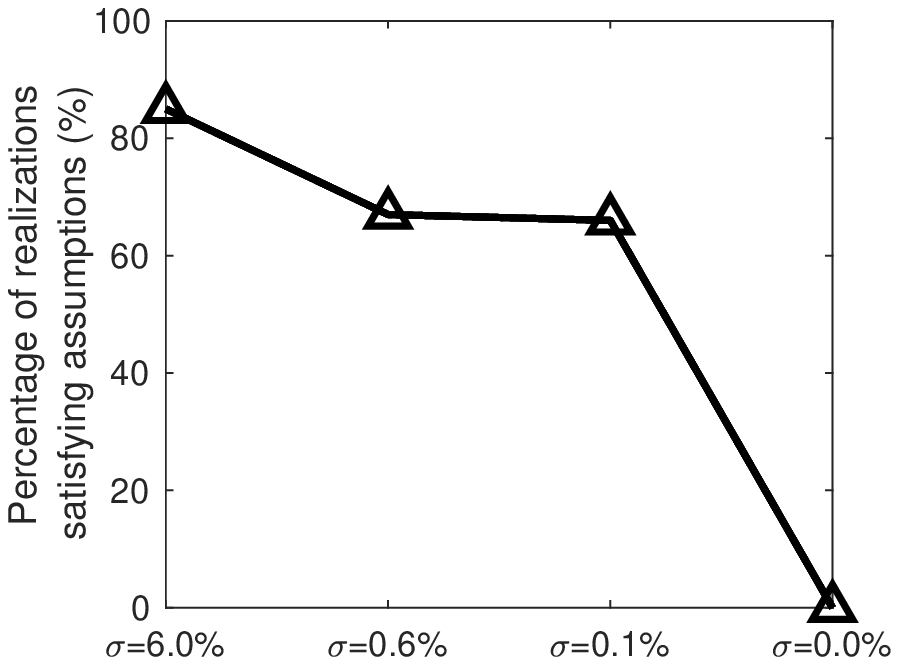}
%\scalefig{0.4}\epsfbox{figs/ISONESetting.eps}
\end{psfrags}
\vspace{-0.5em}\caption{\small  Left: Percentage of realizations satisfied assumptions vs. ramping scenarios from A to H at $\sigma=6\%$. Right: Percentage of realizations satisfied assumptions vs. load forecast error at ramping scenario A.}
 \label{fig:AppendixNewTHMtest}
\end{figure}

It can be observed in the left panel of Fig.~\ref{fig:AppendixNewTHMtest} that 80\% - 90\% realizations satisfied the conditions given in Theorem~\ref{thm:UniformPricingLOC} under ramping scenarios A, B, C, where the system had most binding ramping constraints. From ramping scenarios D to H, binding ramping constraints were gradually relaxed until no binding ramping constraints existed at H, thus less cases satisfied assumptions. The right panel of Fig.~\ref{fig:AppendixNewTHMtest} shows that with larger load forecast error, there were more realizations satisfying the conditions of Theorem~\ref{thm:UniformPricingLOC}.
%Plots of different pricing schemes overlapped in Fig.~\ref{fig:AppendixNewTHMtest}, which shows

We also conducted empirical tests on the  larger ISO-NE case with more practical simulation settings, including network constraints. We observed a higher percentage of the cases satisfying the conditions in Theorem~\ref{thm:UniformPricingLOC}.  Specifically, with  the parameters and load scenarios in the companion paper (Part II)  \cite{Chen&Guo&Tong:20TPS},    99\% - 100\% realizations satisfied the conditions given in Theorem~\ref{thm:UniformPricingLOC} under ramping scenarios A, B, C, D and E.

\subsection{Simulations on the price taker assumption in Theorem~\ref{thm:TLMPBid}}

To mimic a price taking generator, we added generator G4 with marginal cost 35 \$/MWh, generation capacity 1 MW and ramping capacity 0.5 MW under the same parameter settings in Appendix~J. 100 realizations of load scenarios were generated with the same method where $\sigma=6\%$. To validate how frequently the price taker assumption holds for G4, the bidding cost of G4 varied from 34.99 \$/MWh to 35.01 \$/MWh, and the percentage of time intervals was computed when TLMP was not influenced by the changing bid. At ramping setting B with more binding ramping constraints, it's computed that the price-taking assumption held for 64.25\% of the time. And at ramping setting G with less binding ramping constraints, the price-taking assumption held for 81.63\% of the time.
%%At ramping setting B and H, the price-taking assumption held respectively for A  59.54\%, B 64.25\% and 100\% of the time. 

\subsection{Truthful-bidding incentives under R-LMP and R-TLMP}
Under the similar parameter settings as in Example II in Sec~\ref{sec:example}, we  added generator G3 with small generation capacity to mimic a price taking generator and considered the bidding decision process of G3 at $t=1$  as a price taker under the assumption that the true cost of generation is \$28/MWh.  Under the forecasted demand $\hat{\dbf}_{t=1} = (420, 600, 600)$, Table~IV shows the forecasted  $W=2$ window sized rolling-window dispatch of the three generators $\hat{g}_{it}^{\RED}$, the  forecasted rolling-window LMP  $\hat{\pi}_{it}^{\RLMP}$, and the forecasted rolling-window TLMP  $\hat{\pi}_{it}^{\RTLMP}$.  Only the dispatch and pricing decisions at $t=1$ is realized.

\begin{table} [h] \label{tab:Truthful bidding rolling}
\caption{\small Rolling-window economic dispatch, R-LMP, and R-TLMP consider price taker G3. Initial generation  $\gbf[0]={(370, 50, 0)}$. }
\begin{psfrags}
\centerline{
\psfrag{gpp1}[c]{\small $(\hat{g}_{it}^{\RED},\\hat{pi}_t^{\RLMP},\hat{\pi}_{it}^{\RTLMP})$}
\psfrag{t=1}[c]{\small $t=1$}
\psfrag{t=2}[c]{\small $t=2$}
\psfrag{t=3}[c]{\small $t=3$}
\psfrag{d}[c]{ $\hat{\dbf}_t$}
\psfrag{g}[c]{\small $\bar{g}_i$}
\psfrag{c}[c]{\small $c_i$}
\psfrag{r}[c]{\small $\underline{r}_i=\bar{r}_i$}
\scalefig{0.4}\epsfbox{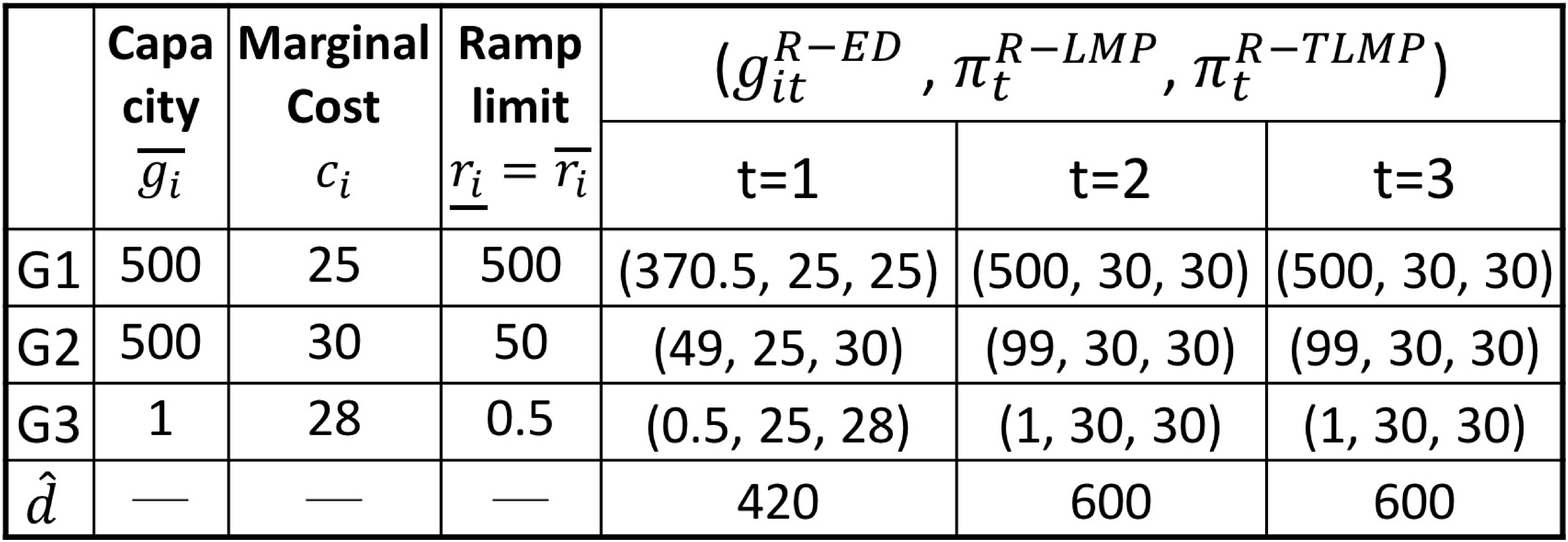}}
\end{psfrags}
\end{table}

\begin{table} [h] \label{tab:Revenue LOC profit}
\caption{\small Ex-ante computation of generation surplus, LOC, and profit of price taker G3.}
\begin{psfrags}
\centerline{
\psfrag{gpp1}[c]{\small $(g_{it}^{\RED},\pi_t^{\RLMP},\pi_{it}^{\RTLMP})$}
\psfrag{t=1}[c]{\small $t=1$}
\psfrag{t=2}[c]{\small $t=2$}
\psfrag{t=3}[c]{\small $t=3$}
\psfrag{d}[c]{ $\hat{\dbf}_t$}
\psfrag{g}[c]{\small $\bar{g}_i$}
\psfrag{c}[c]{\small $c_i$}
\psfrag{r}[c]{\small $\underline{r}_i=\bar{r}_i$}
\scalefig{0.3}\epsfbox{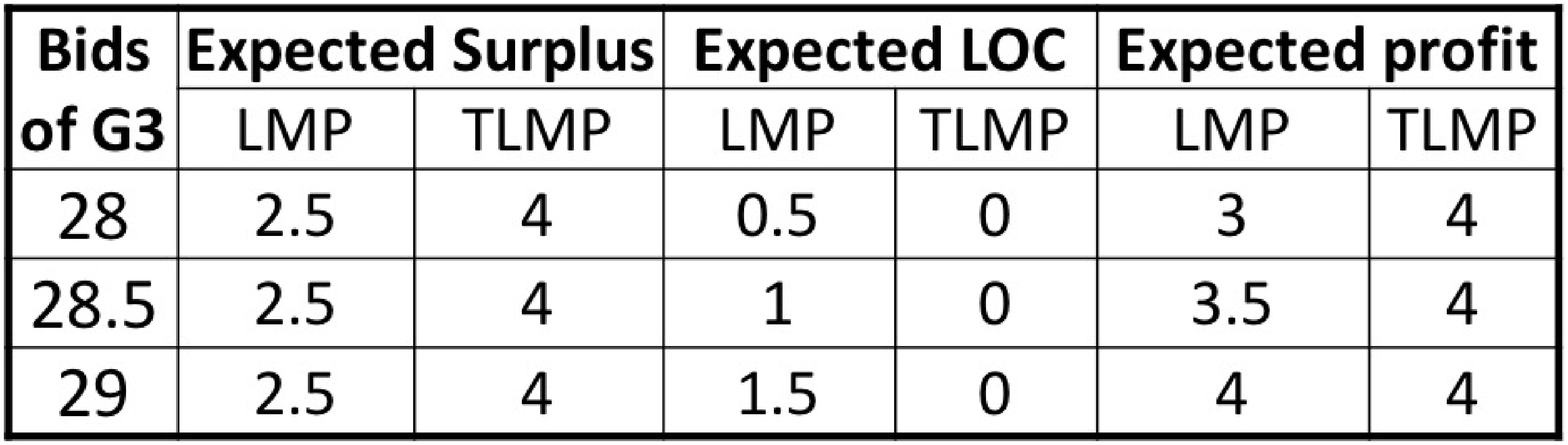}}
\end{psfrags}
\end{table}

Table V shows the {\em expected} surplus, LOC, and total profits of the price-taker G3  under
the rolling-window dispatch and pricing with different bids.  The results showed that, under R-LMP, G3 had higher expected profit when it bid at $\$29$/MWh when true cost is $\$28$/MWh.  Thus there was incentive for the  profit-maximizing price-taker G3 to deviate its bid from the true cost.   Note that the expected generation surpluses were the same under different bids.  Therefore, the gain in profit came entirely from LOC due to untruthful bidding.  In contrast, under R-TLMP, there is no incentive for G3 to bid untruthfully.

%
%\subsection{\tcb{Simulations on the price taker assumption in Theorem~\ref{thm:TLMPBid}}}
%\tcb{
%To mimic a price taking generator, we added generator G4 with marginal cost 35 \$/MW, generation capacity 1 MW and ramping capacity 0.5 MW under the same parameter settings in Appendix~H. 100 realizations of load scenarios were generated with the same method where $\sigma=6\%$. To validate how frequently the price taker assumption holds for G4, the bidding cost of G4 varied from 34.99 \$/MW to 35.01 \$/MW, and the percentage of time intervals was computed when TLMP was not influenced by the changing bid. At ramping setting B with more binding ramping constraints, it's computed that the price-taking assumption held for 64.25\% of the time. And at ramping setting G, the price-taking assumption held for 81.63\% of the time.
%}
%%At ramping setting B and H, the price-taking assumption held respectively for A  59.54\%, B 64.25\% and 100\% of the time. 
%
%\input Appendix_NewTHMTestaV4
%\input Appendix_TBidIncenRLMPV3

% \tcb{
% \input PricetakerBidIncentiveRLMP
% }

% \input Appendix_NewTHM5
\edoc